\documentclass[aps,prb,floats,twocolumn,showpacs]{revtex4}
\usepackage{amssymb}

\usepackage{graphicx,psfig}
\usepackage{dcolumn}
\usepackage{bm}

\input{tcilatex}

\begin{document}

\title{Fronts of spin tunneling in molecular magnets}
\date{\today}
\author{D. A. Garanin}
\affiliation{Physics Department, Lehman College, City University
of New York \\ 250 Bedford Park Boulevard West, Bronx, New York
10468-1589, USA}
\date{\rm 29 April 2009}

\begin{abstract}
Dissipative spin-tunneling transitions at biased resonances in molecular
magnets such as Mn$_{12}$ Ac are controlled by the dipolar field that can
bring the system on and off resonance. It is shown that this leads to spin
relaxation in form of propagating fronts of tunneling, with the dipolar
field adjusting self-consistently to provide a zero bias within the front
core. There are two regimes of the front propagation: laminar and
non-laminar with discontinuous magnetization and dipolar field. In the
laminar regime the speed of the front can exceed that of the magnetic
deflagration, if the transverse field is large enough. Fronts of tunneling
can be initiated by magnetic field sweep near the end of the crystal.
\end{abstract}
\maketitle


\section{Introduction}

Molecular magnets (MM), including their first and mostly studued
representative Mn$_{12}$ Ac, \cite{lis80} have initially attracted attention
as molecules with the effective big spin $S=10$ showing bistability as a
result of a strong uniaxial anisotropy $-DS_{z}^{2}.$ \cite
{sesgatcannov93nat} Resonance spin tunneling manifested in the magnetic
hysteresis loops\cite{frisartejzio96prl,heretal96epl,thoetal96nat} with
steps at the field values $B\approx B_{k}=kD/\left( g\mu _{B}\right) $ , $%
k=0,\pm 1,\pm 2,\ldots $ made molecular magnets a hotspot of research during
more than 10 years.

Crystals of molecular magnets do not show a significant exchange interaction
because the magnetic core of the molecule is screened by organic ligands.
Thus magnetic molecules remain largely superparamagnetic, although MM can
order below 1 K due to dipole-dipole interactions (DDI). \cite
{moretal03prl,garchu08prb}

An important role of the DDI is that the dipolar field created by the spins
is large enough to change the resonance condition for the up and down spins
and thus to strongly influence spin tunneling. Fully ordered spins in an
elongated Mn$_{12}$ Ac crystal create the dipolar field $B^{(D)}\simeq 52.6$
mT at a molecule. \cite{garchu08prb,mchughetal09prb} This becomes comparable
with the resonance width defined by the tunnel splitting $\Delta $ in
transverse magnetic fields above 5 T, for the $k=1$ tunneling resonance. For
smaller transverse fields, $\Delta $ is much smaller and thus the DDI can
completely block the resonant tunneling. The action of the dipolar field is
dynamical and self-consistent since tunneling of spins causes the dipolar
field to change, blocking or allowing resonant transitions.

The role of the DDI in spin tunneling was recognized in Refs.
\onlinecite
{prosta98prl,cucforretadavil99epjb,alofer01prl,feralo03prl,statup04prb,tupstapro04prb,feralo04prb,feralo05prb}
where Monte Carlo simulations were done on the basis of a phenomenological
model involving discrete jumps of the spins through the instantaneous
``tunneling window''. The main purpose of these studies was to explain the $%
\sqrt{t}$ relaxation experimentally observed in Mn$_{12}$ Ac. \cite
{weretal99prl}

It was not understood until recently that DDI in molecular magnets can
result in spatially-inhomogeneous states creating the dipolar field such
that the system is on resonance in some regions of space where spins can
relax, leading to moving fronts. An example is the domain wall in elongated
dipolar-ordered crystals of Mn$_{12}$ Ac at low temperatures. The reduced
dipolar field at $T=0$ in Fig.\ 2 of Ref.\ \onlinecite{garchu08prb} is close
to zero in the region around the center of the domain wall with the width of
the order of the crystal's thickness. It should be mentioned that Mn$_{12}$
Ac remains the only molecular magnet that can be grown in long crystals
required for such kind of phenomena.

Similar effects can take place in the tunneling at biased resonances, $k\geq
1.$ If the external field approaches the resonance by a slow sweep, as was
the case in many experiments, moving walls of tunneling can be created near
the ends of long crystals (where the dipolar bias is smaller) and then
penetrate into their depth with a speed unrelated to the sweep rate. \cite
{garchu09prl} The role of the sweep is only to create an initial state for
the wall of tunneling to start. It was argued that this mechanism can
explain the width of the steps in dynamic hysteresis curves\cite
{frisartejzio96prl,heretal96epl,thoetal96nat} by the time needed for the
wall of tunneling to cross the crystal. Non-uniformity of the magnetization
in Mn$_{12}$ Ac developing during spin tunneling was detected by local
measurements earlier.\cite{avretal05prb}

The walls or self-organized patterns of spin tunneling investigated in Ref.\ %
\onlinecite{garchu09prl} are not exactly fronts because they are lacking the
combined space-time dependence on the argument $z-vt$ only, where $v$ is the
speed of the front. Frozen-in quasiperiodic spatial structures have been
found behind these moving walls. In fact, true smooth fronts of spin
tunneling do exist in the range of the external bias smaller than that in
Ref.\ \onlinecite
{garchu09prl}. Studying these fronts and their transition to the moving
walls with a nonuniformity behind with increasing the bias is the purpose of
this article. It will be shown that there are two regimes.

For the external bias not exceeding a critical value, the true fronts (that
can be called ``laminar'') are realized in which the dipolar field adjusts
to create a resonance in the front region. In the limit of strong dipolar
field (relative to the resonance width) the front speed and the
magnetization behind the front can be calculated analytically and are
independent of the strength of the DDI.

For a larger external bias, the magnetization distribution and thus the
dipolar field in the wall cannot fully adjust to provide the resonance
condition. In this case the wall is moving with a quasiperiodically varying
speed leaving a quasiperiodic state behind. The average wall speed decreases
with the DDI strength quadratically.

The dipolar mechanism of spin tunneling is resembling magnetic deflagration
in Mn$_{12}$ Ac. \cite{suzetal05prl,garchu07prb} Here, instead of the
temperature, the relaxation rate is controlled by the self-consistent
dipolar field bringing the system on or off resonance. Thus, in a sense, one
can call the phenomenon studied here \emph{cold deflagration}. Of course,
the heat release in the course of the cold deflagration can give rise to the
regular deflagration, especially for high resonances $k$ and well thermally
isolated crystals. In this case the two kinds of deflagration can compete.

The rest of the article is organized as follows. In Sec. \ref{Sec-tun-relax}
the dynamics of spin tunneling between the metastable ground state and a
resonant excited state on the other side of the barrier is considered. The
simplified overdamped equations of motion are obtained in the case of the
tunnel splitting frequency $\Delta /\hbar $ smaller than the damping of the
excited state $\Gamma .$ It is further argued that in the presence of
disorder that spreads resonances one can use overdamped equations in a
generalized form also for larger $\Delta .$ In Sec. \ref{Sec-dipolar} the
dipolar field created by a wall of magnetization is calculated for the
cylindrical and ribbon geometries. In Sec. \ref{Sec-cold-deflagr-eqs} the
full system of cold deflagration equations is written and transformed into
dimensionless form. In Sec. \ref{Sec-strong-ED} the limit of strong DDI is
studied and analytical expressions for the residual magnetization behind the
front and the front speed are obtained. Sec. \ref{Sec-numerical} provides
the results of numerical calculations in both regimes of the wall
propagation.

\section{Spin tunneling and relaxation}

\label{Sec-tun-relax}

We will be using the generic giant-spin model of molecular magnets with the
Hamiltonian
\begin{equation}
\hat{H}=-DS_{z}^{2}-g\mu _{B}B_{z}S_{z}-g\mu _{B}B_{x}S_{x}+\ldots ,
\label{MMHam}
\end{equation}
where $D$ is the uniaxial anosotropy and
\begin{equation}
\mathbf{B=B}_{\mathrm{ext}}+\mathbf{B}^{(D)}  \label{Bdef}
\end{equation}
is the total magnetic field, including the external and dipolar fields.
Suppressed terms in the Hamiltonian can include the biaxial and fourth-order
anisotropy that can make a contribution into the tunnel splitting $\Delta $
of the resonant spin up- and down states. Since the most interesting
situation arises in the case of a large $\Delta $ that can only be created
by a strong transverse field, the dropped terms will not be needed. For $%
B_{x}=0$ the exact quantum states of $\hat{H}$ are $\left| m\right\rangle $
with $-S\leq m\leq S,$ their energies being $\varepsilon _{m}=-Dm^{2}-g\mu
_{B}B_{z}m.$ The resonance condition $\varepsilon _{m}=\varepsilon
_{m^{\prime }}$ between \emph{all} states $\left| m\right\rangle $ on the
left side of the barrier ($m<0)$ and $\left| m^{\prime }\right\rangle $ on
the right side of the barrier ($m^{\prime }=-m-k$) is satisfied for the
resonance fields
\begin{equation}
B_{z}=B_{k},\qquad B_{k}=kD/\left( g\mu _{B}\right) ,\qquad k=0,1,\ldots
\label{BkDef}
\end{equation}
This resonance condition turns out to be independent of the transverse field.

Application of the transverse field leads to the two effects. First, each
state $\left| m\right\rangle $ hybridyzes with neighbouring states within
the same well forming the state that can be denoted as $\left| \psi
_{m}\right\rangle .$ Physically this corresponds to spin canting in the
direction of the transverse field. Second, the states $\left| \psi
_{m}\right\rangle $ on different sides of the barrier hybridize because of
the resonance spin tunneling near $B_{z}=B_{k}.$ Of course, one can speak of
the states $\left| \psi _{m}\right\rangle $ for not too strong transverse
field, so that there still are low-lying states well localized within one of
the wells. The states $\left| \psi _{m}\right\rangle $ provide a basis for a
simplified treatment of spin tunneling and relaxation near resonances that
otherwise has to be considered within the density-matrix formalism. \cite
{garchu97prb,gar08-DME}

Consider the metastable ground state of a molecular magnet, $\left| \psi
_{m}\right\rangle =\left| \psi _{-S}\right\rangle ,$ near a tunneling
resonance with an excited state $\left| \psi _{m^{\prime }}\right\rangle $
on the right side of the barrier, Fig. 1 of Ref. \onlinecite{garchu09prl}.
The dynamics of tunneling at low temperatures is described by the subset of
the density matrix equation (DME) taking into account only these two levels.
The level $\left| \psi _{m^{\prime }}\right\rangle $ can decay into
lower-lying levels within the same well with rate $\Gamma _{m^{\prime }}$.
Since there are no incoming relaxation processes for the state $\left| \psi
_{m^{\prime }}\right\rangle $ at low temperatures, the DME can be simplified
to the form of the damped Schr\"{o}dinger equation
\begin{eqnarray}
\dot{c}_{-S} &=&-\frac{i}{2}\frac{\Delta }{\hbar }c_{m^{\prime }}  \nonumber
\\
\dot{c}_{m^{\prime }} &=&\left( \frac{iW}{\hbar }-\frac{1}{2}\Gamma
_{m^{\prime }}\right) c_{m^{\prime }}-\frac{i}{2}\frac{\Delta }{\hbar }%
c_{-S}.  \label{DME-DampedSchr}
\end{eqnarray}
Here $\Delta $ is the tunnel splitting and $W\equiv \varepsilon
_{m}-\varepsilon _{m^{\prime }}$ is the energy bias between the two levels,
\begin{equation}
W=\left( S+m^{\prime }\right) g\mu _{B}\left( B_{z}-B_{k}+B_{z}^{(D)}\right)
\equiv W_{\mathrm{ext}}+W^{(D)}.  \label{WiDef}
\end{equation}
In fact, here one should use the values of $S$ and $m^{\prime }$ corrected
for spin canting. Since $\left| \psi _{-S}\right\rangle $ is the lowest
state in the left well, it cannot decay. The numbers of particles in the
states are defined by
\begin{equation}
n_{-S}=\left| c_{-S}\right| ^{2},\qquad n_{m^{\prime }}=\left| c_{m^{\prime
}}\right| ^{2}  \label{phodiagDef}
\end{equation}
etc. The spin polarization in our low-temperature tunneling process is given
by $\left\langle S_{z}\right\rangle =-Sn_{-S}+\sum_{m=m^{\prime }}^{S}mn_{m}.
$ As the states with $m=m^{\prime }+1,\ldots ,S-1$ decay faster than $\left|
\psi _{m^{\prime }}\right\rangle ,$ their contribution can be neglected.
Then for the normalized spin-average variable
\begin{equation}
\sigma _{z}\equiv \left\langle S_{z}\right\rangle /S,  \label{sigmazDef}
\end{equation}
one obtains
\begin{equation}
\sigma _{z}=1-2n_{-S}-(1-m^{\prime }/S)n_{m^{\prime }}.  \label{sigmazAvr}
\end{equation}

In the overdamped case $\Gamma _{m^{\prime }}\gg \Delta /\hbar $ the
variable $c_{m^{\prime }}$ in Eq.\ (\ref{DME-DampedSchr}) adiabatically
adjusts to the instantaneous value of $c_{-S}.$ Setting $\dot{c}_{m^{\prime
}}=0$ in the second of these equations one obtains
\begin{equation}
c_{m^{\prime }}=\frac{\Delta }{2\hbar }\frac{c_{-S}}{W/\hbar +i\Gamma
_{m^{\prime }}/2}.
\end{equation}
Inserting it into the first of equations (\ref{DME-DampedSchr}) one obtains
\begin{equation}
\dot{c}_{-S}=-\frac{i\Delta ^{2}}{4\hbar ^{2}}\frac{c_{-S}}{W/\hbar +i\Gamma
_{m^{\prime }}/2}.
\end{equation}
With the help of Eq.\ (\ref{phodiagDef}), one obtains the equation for the
metastable population $n_{-S}$ in the form
\begin{equation}
\dot{n}=-\Gamma n,  \label{dotrhoEq}
\end{equation}
where the subscript $-S$ has been dropped for transparency and the
dissipative tunneling rate $\Gamma $ is given by \cite{garchu97prb}
\begin{equation}
\Gamma =\frac{\Delta ^{2}}{2\hbar ^{2}}\frac{\Gamma _{m^{\prime }}/2}{\left(
W/\hbar \right) ^{2}+\left( \Gamma _{m^{\prime }}/2\right) ^{2}}.
\label{GammaDef}
\end{equation}
In the overdamped limit one has $n_{m^{\prime }}\ll n_{-S}\equiv n,$ so that
Eq.\ (\ref{sigmazAvr}) simplifies to
\begin{equation}
\sigma _{z}=1-2n.  \label{sigmazAvrOverd}
\end{equation}

The decay rate $\Gamma _{m^{\prime }}$ is mainly due to the relaxation
between the adjacent energy levels in the right well, $\Gamma _{m^{\prime
}}=\Gamma _{m^{\prime },m^{\prime }+1}$, see Eq.\ (A9) of Ref.\
\onlinecite
{chugarsch05prb} or Eq.\ (294) of Ref.\ \onlinecite{gar08-DME}. For $k=1$
and thus $m^{\prime }=9$ one has $\Gamma _{m^{\prime }}\approx 10^{7}$ s$%
^{-1},$ while $\Delta /\hbar $ reaches a comparable value in the transverse
field above 3 T. At higher transverse fields the tunneling dynamics should
be underdamped. Resonances with higher $k$ have larger splitting $\Delta $
and become underdamped in smaller transverse fields.

In the case of underdamped resonances, $\Delta /\hbar \gtrsim \Gamma
_{m^{\prime }},$ the rate of dissipative spin tunneling can be described by
the integral relaxation time $\tau _{\mathrm{int}}$ resulting in the
effective rate\cite{gar08-DME}
\begin{equation}
\Gamma =\frac{1}{\tau _{\mathrm{int}}}=\frac{\Delta ^{2}}{2\hbar ^{2}}\frac{%
\Gamma _{m^{\prime }}/2}{\Omega ^{2}+\left( \Gamma _{m^{\prime }}/2\right)
^{2}},  \label{Gammatauint}
\end{equation}
where
\begin{equation}
\left( \hbar \Omega \right) ^{2}\equiv W^{2}+\frac{1}{4}\left( 1+\frac{%
S-m^{\prime }}{2S}\right) \Delta ^{2}.  \label{OmegaDef}
\end{equation}
One can see that in the underdamped case the width of the Lorentzian becomes
of the order of $\left( \Delta /2\right) /\hbar $, compared to $\Gamma
_{m^{\prime }}/2$ in the overdamped case. Although Eq.\ (\ref{dotrhoEq})
with $\Gamma $ given by Eq.\ (\ref{Gammatauint}) does not accurately
describe the oscillating dynamics of the system in the underdamped case, in
particular the Landau-Zener effect, it will be used below as an
approximation for the many-body problem with coupling via the dipolar field
in both overdamped and underdamped cases. A more rigorous approach based on
Eq.\ (\ref{DME-DampedSchr}) requires much more computer time because of fast
oscillations. On the other hand, oscillations at tunneling resonances have
never been experimentally observed in MM because of the resonance spread as
a result of ligand disorder and other factors. For the low-bias resonances
such as $k=1$ and thus $m^{\prime }=S-1$ the contribution of $n_{m^{\prime }}
$ in Eq.\ (\ref{sigmazAvr}) can be neglected, thus Eq.\ (\ref{sigmazAvrOverd}%
) will be used in all cases.

\section{Dipolar field}

\label{Sec-dipolar}

The dipolar field and ensuing dipolar bias of tunneling resonances in
crystals of molecular magnets have been discussed in detailes in Ref.\
\onlinecite
{garchu08prb}, so that only a short summary with necessary changes will be
provided below.

The $z$ component of dipolar field at site $i$ (i.e., at a particular
magnetic molecule)$,$ created by \ molecular spins polarized along the $z$
axis is given by
\begin{equation}
B_{i,z}^{(D)}=\frac{Sg\mu _{B}}{v_{0}}D_{i,zz},\qquad D_{i,zz}\equiv
\sum_{j}\phi _{ij}\sigma _{jz},  \label{BDiz}
\end{equation}
where $v_{0}$ is the unit-cell volume and
\begin{equation}
\phi _{ij}=v_{0}\frac{3\left( \mathbf{e}_{z}\cdot \mathbf{n}_{ij}\right)
^{2}-1}{r_{ij}^{3}},\qquad \mathbf{n}_{ij}\equiv \frac{\mathbf{r}_{ij}}{%
r_{ij}}.  \label{phizijDef}
\end{equation}
Inside a uniformly magnetized ellipsoid the dipolar field is uniform and one
has $D_{zz}=\bar{D}_{zz}\sigma _{z},$ where
\begin{equation}
\bar{D}_{zz}=\bar{D}_{zz}^{(\mathrm{sph})}+4\pi \nu \left(
1/3-n^{(z)}\right) ,  \label{DzzEllipsoid1}
\end{equation}
$\nu $ is the number of magnetic molecules per unit cell and $n^{(z)}=0,$ $%
1/3,$ and 1 for a cylinder, sphere, and disc, respectively. $\bar{D}_{zz}^{(%
\mathrm{sph})}$ depends on the lattice structure. For Mn$_{12}$ Ac lattice
summation yields $\bar{D}_{zz}^{(\mathrm{sph})}=2.155$ that results in $\bar{%
D}_{zz}^{(\mathrm{cyl})}=10.53$ for a cylinder. Then Eq.\ (\ref{BDiz})
yields the dipolar field $B_{z}^{(\mathrm{D})}\simeq 52.6$ mT in an
elongated sample that was also obtained experimentally.\cite{mchughetal09prb}

The dipolar bias $W^{(D)}$\ in Eq.\ (\ref{WiDef}) can be written in the form
\begin{equation}
W^{(D)}=\left( 1+\frac{m^{\prime }}{S}\right) E_{D}D_{zz},  \label{WiDDef}
\end{equation}
where
\begin{equation}
E_{D}\equiv \left( Sg\mu _{B}\right) ^{2}/v_{0}  \label{EDDef}
\end{equation}
is the dipolar energy, $E_{D}/k_{B}=0.0671$ K for Mn$_{12}$ Ac. Since the
dipolar field depends on the magnetization and its values in an elongated
crystal can change between $-52.6$ mT and $52.6$ mT, one can conclude that,
according to Eq.\ (\ref{WiDef}), the resonance condition $W=0$ can be, in
principle, satisfied in the dipolar tunneling window around the resonance $%
-52.6$ mT $\leq B_{z}-B_{zk}\leq 52.6$ mT. This dipolar window is much
smaller than the distance between the two tunneling resonances that is about
0.5 T. Practically, for a negative external bias $B_{z}-B_{zk}$ the
relaxation is hindered by the causality: To produce a positive dipolar field
that would balance the negative external bias, spins should already be on
the right side of the barrier.

For a cylinder of length $L$ and radius $R$ with the symmetry axis $z$ along
the easy axis, magnetized with $\sigma _{z}=$ $\sigma _{z}(z),$ the reduced $%
z$-field along the symmetry axis has the form
\begin{equation}
D_{zz}(z)=\int_{-L/2}^{L/2}dz^{\prime }\frac{2\pi \nu R^{2}\sigma
_{z}(z^{\prime })}{\left[ \left( z^{\prime }-z\right) ^{2}+R^{2}\right]
^{3/2}}-k\sigma _{z}(z),  \label{DzzCylinder}
\end{equation}
where
\begin{equation}
k\equiv 8\pi \nu /3-\bar{D}_{zz}^{(\mathrm{sph})}=4\pi \nu -\bar{D}_{zz}^{(%
\mathrm{cyl})}>0,  \label{kDef}
\end{equation}
$k=14.6$ for Mn$_{12}$ Ac. In particular, for a long uniformly polarized
cylinder one obtains
\begin{equation}
D_{zz}(z)=\frac{2\pi \nu z}{\sqrt{z^{2}+R^{2}}}\sigma _{z}+\left( \bar{D}%
_{zz}^{(\mathrm{sph})}-2\pi \nu /3\right) \sigma _{z},
\label{DzzCylinderLong}
\end{equation}
At the left end of the long cylinder, $z=0,$ one has $D_{zz}=\left( \bar{D}%
_{zz}^{(\mathrm{sph})}-2\pi \nu /3\right) \sigma _{z}.$ For Mn$_{12}$ Ac one
obtains $D_{zz}=-2.03\sigma _{z}$, having the sign opposite to that of the
field in the depth, $\bar{D}_{zz}^{(\mathrm{cyl})}\sigma _{z}$. This means
that a homogeneously magnetized state in the resonance external field $%
B_{z}=B_{zk}$ may be unstable with respect to spin tunneling beginning in
the vicinity of the ends of the cylinder, as at some point near the end the
resonance condition $D_{zz}(z)=0$ is satisfied. To the contrary, dipolar
field in the depth of a uniformly magnetized cylinder provides the dipolar
bias that puts the system off resonance and makes the transition rate $%
\Gamma $ very small. If the external field $B_{z}$ is swept in the positive
direction towards the resonance, spin tunneling begins near the ends of the
crystal and then it propagates inside the crystal as a moving wall of
tunneling, as the dipolar field changes self-consistently. \cite{garchu09prl}

One can also calculate the dipolar field on the $z$ symmetry axis of a slab
of length $L$ and sides $a$ and $b.$ In the case $a\sim b$ the results are
similar to those for the cylinder above. For a thick slab (a ribbon) with $%
a\ll b$ one obtains
\begin{equation}
D_{zz}(z)=\int_{-L/2}^{L/2}dz^{\prime }\frac{2a\nu \sigma _{z}(z^{\prime })}{%
\left( z^{\prime }-z\right) ^{2}+(a/2)^{2}}-k\sigma _{z}(z)
\label{DzzThickSlab}
\end{equation}
that has the kernel less localized than that of Eq.\ (\ref{DzzCylinder}).

\section{Cold deflagration equations}

\label{Sec-cold-deflagr-eqs}

The phenomenon of cold deflagration is described by a collection of
equations (\ref{dotrhoEq}) for every magnetic molecule in the crystal, with
the dipolar field controlling transitions being determined by the
instantaneous non-uniform magnetization. As the full three-dimensional
problem with a long-range interaction requires too much computer power, here
the one-dimensional approximation will be made. The magnetization is
considered as a function of the coordinate $z$ only$,$ i.e., the
deflagration fronts are flat, and the dipolar field is taken along the
symmetry axis as in Eq.\ (\ref{DzzCylinder}) that will be used below. Of
course, the dipolar field away from the symmetry axis is different, that
will result in non-flat fronts. However, to avoid complications in
demonstrating the basic phenomenon, these effects will be ignored here.

It is convenient to introduce the dimensionless time $\widetilde{t}$ and
coordinate $\widetilde{z}$ as
\begin{equation}
\widetilde{t}\equiv \Gamma _{\mathrm{res}}t,\qquad \widetilde{z}\equiv z/R,
\label{tauuDef}
\end{equation}
where $\Gamma _{\mathrm{res}}$ is the resonance relaxation rate following
from Eq.\ (\ref{Gammatauint}) with $\left( S-m^{\prime }\right) /\left(
2S\right) $ neglected,
\begin{equation}
\Gamma _{\mathrm{res}}=\frac{\Delta ^{2}}{\hbar ^{2}}\frac{\Gamma
_{m^{\prime }}}{\left( \Delta /\hbar \right) ^{2}+\Gamma _{m^{\prime }}^{2}}.
\label{GammaResDef}
\end{equation}
Then Eq.\ (\ref{dotrhoEq}) becomes
\begin{equation}
\frac{d}{d\widetilde{t}}\,n(\widetilde{z},\widetilde{t})=-F(\widetilde{z},%
\widetilde{t})n(\widetilde{z},\widetilde{t}),  \label{nztEq}
\end{equation}
where $F$ contains integral dependence on $n(\widetilde{z},\widetilde{t})$
via $D_{zz}(\widetilde{z},\widetilde{t})$,
\begin{equation}
F(\widetilde{z},\widetilde{t})=\frac{1}{1+4\widetilde{E}_{D}^{2}\widetilde{W}%
^{2}(\widetilde{z},\widetilde{t})}\,,\qquad \widetilde{E}_{D}\equiv \frac{%
2E_{D}}{\sqrt{\Delta ^{2}+\hbar ^{2}\Gamma _{m^{\prime }}^{2}}}\,.
\label{FDef}
\end{equation}
The dimensionless bias $\widetilde{W}=W/\left( 2E_{D}\right) ,$ with $W$
defined by Eq.\ (\ref{WiDef}), has the form
\begin{equation}
\widetilde{W}(\widetilde{z},\widetilde{t})=\widetilde{W}_{\mathrm{ext}}+%
\widetilde{W}^{(D)}(\widetilde{z},\widetilde{t})=\widetilde{W}_{\mathrm{ext}%
}+D_{zz}(\widetilde{z},\widetilde{t}),  \label{WtilfDDe}
\end{equation}
where
\begin{equation}
\widetilde{W}_{\mathrm{ext}}=\frac{Sg\mu _{B}}{E_{D}}\left(
B_{z}-B_{k}\right)  \label{WtilExtFinal}
\end{equation}
and $D_{zz}(\widetilde{z},\widetilde{t})$ defined by Eq.\ (\ref{DzzCylinder}%
) can be rewritten in the form
\begin{equation}
D_{zz}(\widetilde{z},\widetilde{t})=\int_{-\widetilde{L}/2}^{\widetilde{L}%
/2}d\widetilde{z}^{\prime }\frac{2\pi \nu \sigma _{z}(\widetilde{z}^{\prime
},\widetilde{t})}{\left[ \left( \widetilde{z}^{\prime }-\widetilde{z}\right)
^{2}+1\right] ^{3/2}}-k\sigma _{z}(\widetilde{z},\widetilde{t})\,,
\label{WzzCylinder}
\end{equation}
where $\sigma _{z}(\widetilde{z},\widetilde{t})=1-2n_{-S}(\widetilde{z},%
\widetilde{t})$ and $\widetilde{L}\equiv L/R.$

For a long sample, $\widetilde{L}\rightarrow \infty ,$ one can seek for a
solution of Eq.\ (\ref{nztEq}) in the form of a moving front depending on
the combined argument $\xi \equiv z-vt,$ where $v$ is the front speed. In
reduced units one has $\widetilde{\xi }\equiv \widetilde{z}-v^{\ast }%
\widetilde{t},$ where the relation between the real and reduced front speeds
has the form
\begin{equation}
v=v^{\ast }\Gamma _{\mathrm{res}}R.  \label{vviavstar}
\end{equation}
Eq.\ (\ref{nztEq}) for the front becomes
\begin{equation}
v^{\ast }\frac{dn}{d\widetilde{\xi }}=F(\widetilde{\xi })n,  \label{dndzEq}
\end{equation}
where $F(\widetilde{\xi })=\left( 1+4\widetilde{E}_{D}^{2}\widetilde{W}^{2}(%
\widetilde{\xi })\right) ^{-1}$ contains
\begin{equation}
D_{zz}(\widetilde{\xi })=\int_{-\infty }^{\infty }d\widetilde{\xi }^{\prime }%
\frac{2\pi \nu \sigma _{z}(\widetilde{\xi }^{\prime })}{\left[ \left(
\widetilde{\xi }^{\prime }-\widetilde{\xi }\right) ^{2}+1\right] ^{3/2}}%
-k\sigma _{z}(\widetilde{\xi })\,.  \label{Dzzvsxi}
\end{equation}
Eq.\ (\ref{vviavstar}) makes the dependence of $v\ $on $R$ and $\Gamma _{%
\mathrm{res}}$ obvious. However, there are nontrivial parameters $\widetilde{%
E}_{D}$ and $\widetilde{W}_{\mathrm{ext}}$ \ that enter $v^{\ast }.$

For a ribbon one can introduce $\widetilde{z}\equiv 2z/a$ and replace Eqs.\ (%
\ref{WzzCylinder}) and (\ref{Dzzvsxi}) by the corresponding expressions
following from Eq.\ (\ref{DzzThickSlab}).

\begin{figure}[tbp]
\includegraphics[angle=-90,width=8cm]{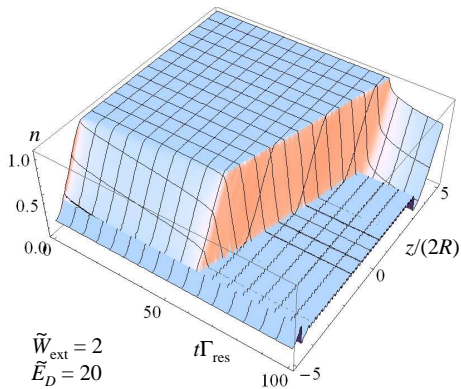}
\caption{Propagating front of cold deflagration for $\widetilde{W}_{\mathrm{%
ext}}=2$ and $\widetilde{E}_{D}=20$. Cold deflagration starts after some
ignition time that depends on the initial condition.}
\label{Fig-n_3d_ED=20_Wext=2}
\end{figure}

\begin{figure}[tbp]
\includegraphics[angle=-90,width=8cm]{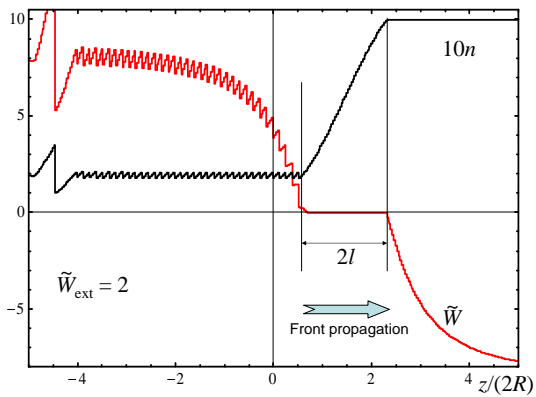}
\caption{Spatial profiles of the metastable population $n$ and the reduced
bias $\widetilde{W}$ in the front for $\widetilde{W}_{\mathrm{ext}}=2$ and $%
\widetilde{E}_{D}=20.$ Everywhere in the front the system is near the
resonance, $\widetilde{W}\approx 0.$ At this value of the bias periodic
structures behind the front begin to emerge. }
\label{Fig-n_W_profiles_ED=20_Wext=2}
\end{figure}

The cold deflagration equations can be solved by discretization that reduces
them to a system of nonlinear ordinary first-order differential equations.
The front of tunneling for $\widetilde{E}_{D}=20$ and $\widetilde{W}_{%
\mathrm{ext}}=2$ is shown in Figs.\ \ref{Fig-n_3d_ED=20_Wext=2} and \ref
{Fig-n_W_profiles_ED=20_Wext=2} while more details will be given in Sec. \ref
{Sec-numerical} and \ref{Sec-Sweep}.

\section{The limit of strong dipolar field}

\label{Sec-strong-ED}

Unless a strong transverse field is applied, the numerical value of $%
\widetilde{E}_{D}$ of Eq.\ (\ref{FDef}) for Mn$_{12}$ Ac is large. In
particular, in the overdamped limit $\widetilde{E}_{D}\cong E_{D}/\left(
\hbar \Gamma _{m^{\prime }}\right) \approx 10^{3}$ for $m^{\prime }=S-1.$ In
this case one could think that $F$ is negligibly small everywhere except for
a very close vicinity of the resonance, so that the total relaxation and
thus the speed of the front are very small. However, as the numerical
solution shows, the system finds the way to relax faster by forming a front
region of a width $l\sim R$ where $\widetilde{W}$ \ is small and $F$ is of
order one. In this extended region resonant tunneling transitions take
place. Beyond the front region $\widetilde{W}$ deviates from zero and $F$
becomes negligibly small. As a result, $n(\xi )$ changes practically only
within the front core.

Basing on these insights, one can construct a perturbative expansion in
powers of $1/\widetilde{E}_{D}$ and show that the solution $n(\widetilde{\xi
})$ and the front speed $v^{\ast }$\ become independent of $\widetilde{E}%
_{D} $ for $\widetilde{E}_{D}\gg 1.$ One can search for $\sigma _{z}(%
\widetilde{\xi })$ in the form
\begin{equation}
\sigma _{z}(\widetilde{\xi })\cong \sigma _{z}^{(0)}(\widetilde{\xi }%
)+\sigma _{z}^{(1)}(\widetilde{\xi })/\widetilde{E}_{D},  \label{nuExpansion}
\end{equation}
and similarly for $n(\widetilde{\xi })=\left[ 1-\sigma _{z}(\widetilde{\xi })%
\right] /2.$ The term $\sigma _{z}^{(0)}(\widetilde{\xi })$ is defined by
the condition that within the front region $-\widetilde{l}\leq \widetilde{%
\xi }\leq \widetilde{l},$ with the width $\widetilde{l}$ to be determined
self-consistently, the contribution of $\sigma _{z}^{(0)}(\widetilde{\xi })$
into the bias $\widetilde{W}$ is zero. If this is fulfilled, the term $4%
\widetilde{E}_{D}^{2}\widetilde{W}^{2}$ in the denominator of Eq.\ (\ref
{FDef}) is of order one due to the correction $\sigma _{z}^{(1)}(\widetilde{%
\xi }), $ so that in the front region $F$ is of order one. In the region
before the front, $\widetilde{l}<\widetilde{\xi },$ one has $n^{(0)}(%
\widetilde{\xi })=1 $ and $\sigma _{z}^{(0)}(\widetilde{\xi })=\sigma
_{zi}=-1.$ Everywhere behind the front, $\widetilde{\xi }<-\widetilde{l},$
one has final values $n^{(0)}(\widetilde{\xi })=n_{f}$ \ and $\sigma
_{z}^{(0)}(\widetilde{\xi })=\sigma _{zf}$ \ that are to be determined. In
the front region the condition $\widetilde{W}=0$ with Eq.\ (\ref{WtilfDDe})
yields the integral equation
\begin{equation}
\widetilde{W}^{(0)}(\widetilde{\xi })=0,\qquad -\widetilde{l}\leq \widetilde{%
\xi }\leq \widetilde{l},  \label{ZeroApprEq}
\end{equation}
where
\begin{eqnarray}
&&\widetilde{W}^{(0)}(\widetilde{\xi })=\widetilde{W}_{\mathrm{ext}}+\int_{-%
\widetilde{l}}^{\widetilde{l}}d\widetilde{\xi }^{\prime }\frac{2\pi \nu
\sigma _{z}^{(0)}(\widetilde{\xi }^{\prime })}{\left[ \left( \widetilde{\xi }%
^{\prime }-\widetilde{\xi }\right) ^{2}+1\right] ^{3/2}}  \nonumber \\
&&{}-k\sigma _{z}^{(0)}(\widetilde{\xi })+2\pi \nu \left[ \sigma _{zi}\left(
1+\frac{\widetilde{\xi }-\widetilde{l}}{\sqrt{(\widetilde{\xi }-\widetilde{l}%
)^{2}+1^{2}}}\right) \right.  \nonumber \\
&&+\left. \sigma _{zf}\left( 1-\frac{\widetilde{\xi }+\widetilde{l}}{\sqrt{(%
\widetilde{\xi }+\widetilde{l})^{2}+1^{2}}}\right) \right] .
\label{ZeroApproxW}
\end{eqnarray}
This equation determines the zero-order profile $\sigma _{z}^{(0)}(%
\widetilde{\xi }),$ including $\widetilde{l}$ and $\sigma _{zf}.$

Eq.\ (\ref{ZeroApprEq}) can be solved numerically by discretizing the
integral using $N+1$ equidistant points within the interval $\left( -%
\widetilde{l},\widetilde{l}\right) $ given by $\widetilde{\xi }_{i}=-$ $%
\widetilde{l}+2\widetilde{l}i/N,$ $i=0,1,\ldots ,N.$ The value at the right
end of the interval is fixed by the boundary condition $\sigma _{z}^{(0)}(%
\widetilde{\xi }_{N})=\sigma _{zi}=-1.$ Thus there are total $N+1$ unknowns
including $\widetilde{l},$ that can be found by solving the system of $N+1$
equations $\widetilde{W}^{(0)}(\widetilde{\xi }_{i})=0$ with $i=0,1,\ldots
,N.$ Note that this system of equations is nonlinear because of $\widetilde{l%
}.$ In this way one finds the zero-order magnetization profile in the front
and the magnetization behind the front $\sigma _{zf}$ for any $\widetilde{W}%
_{\mathrm{ext}}>0.$ In particular, for Mn$_{12}$ Ac one obtains $\widetilde{l%
}=0.848.$

On the other hand, one can find important analytical results if one searches
for the solution in the form
\begin{equation}
\sigma _{z}^{(0)}(\widetilde{\xi })=\frac{\sigma _{zf}+\sigma _{zi}}{2}-%
\frac{\sigma _{zf}-\sigma _{zi}}{2}f(\widetilde{\xi }),  \label{fuDef}
\end{equation}
where $f(\pm \widetilde{l})=\pm 1.$ Substituting this into Eq.\ (\ref
{ZeroApproxW}) one obtains the equation for $f(\widetilde{\xi })$
\begin{eqnarray}
&&0=\widetilde{W}_{\mathrm{ext}}-\frac{\sigma _{zf}-\sigma _{zi}}{2}\int_{-%
\widetilde{l}}^{\widetilde{l}}d\widetilde{\xi }^{\prime }\frac{2\pi \nu f(%
\widetilde{\xi }^{\prime })}{\left[ \left( \widetilde{\xi }^{\prime }-%
\widetilde{\xi }\right) ^{2}+1\right] ^{3/2}}  \nonumber \\
&&{}+k\frac{\sigma _{zf}-\sigma _{zi}}{2}f(\widetilde{\xi })+\bar{D}_{zz}^{(%
\mathrm{cyl})}\frac{\sigma _{zf}+\sigma _{zi}}{2}  \nonumber \\
&&-\pi \nu \left( \sigma _{zf}-\sigma _{zi}\right) \left( \frac{\widetilde{%
\xi }+\widetilde{l}}{\sqrt{(\widetilde{\xi }+\widetilde{l})^{2}+1^{2}}}+%
\frac{\widetilde{\xi }-\widetilde{l}}{\sqrt{(\widetilde{\xi }-\widetilde{l}%
)^{2}+1^{2}}}\right) ,  \nonumber \\
&&  \label{ZeroApprEqf}
\end{eqnarray}
where $\bar{D}_{zz}^{(\mathrm{cyl})}=4\pi \nu -k.$ One can see that there
are even and odd terms in $\widetilde{\xi }$ in this equation and $f(%
\widetilde{\xi })$ is odd. The even and odd parts of this equation should
turn to zero independently of each other. For the even part one obtains the
equation
\begin{equation}
0=\widetilde{W}_{\mathrm{ext}}+\bar{D}_{zz}^{(\mathrm{cyl})}\frac{\sigma
_{zf}+\sigma _{zi}}{2}
\end{equation}
that with $\sigma _{zi}=-1$ yields $\ \sigma _{zf}=1-2\widetilde{W}_{\mathrm{%
ext}}/\bar{D}_{zz}^{(\mathrm{cyl})}$ and
\begin{equation}
n_{f}=\frac{1-\sigma _{zf}}{2}=\frac{\widetilde{W}_{\mathrm{ext}}}{\bar{D}%
_{zz}^{(\mathrm{cyl})}}  \label{nFinal}
\end{equation}
for the fraction of metastable molecules behind the front. Note that since $%
0\leq n_{f}\leq 1,$ this solution only exists for
\begin{equation}
0\leq \widetilde{W}_{\mathrm{ext}}\leq \bar{D}_{zz}^{(\mathrm{cyl})}.
\label{WextRange}
\end{equation}

The odd part of Eq.\ (\ref{ZeroApprEqf}) yields the equation
\begin{eqnarray}
&&-\int_{-\widetilde{l}}^{\widetilde{l}}d\widetilde{\xi }^{\prime }\frac{f(%
\widetilde{\xi }^{\prime })}{\left[ \left( \widetilde{\xi }^{\prime }-%
\widetilde{\xi }\right) ^{2}+1\right] ^{3/2}}+\frac{k}{2\pi \nu }f(%
\widetilde{\xi })  \nonumber \\
&=&\frac{\widetilde{\xi }+\widetilde{l}}{\sqrt{(\widetilde{\xi }+\widetilde{l%
})^{2}+1^{2}}}+\frac{\widetilde{\xi }-\widetilde{l}}{\sqrt{(\widetilde{\xi }-%
\widetilde{l})^{2}+1^{2}}}  \label{fuEq}
\end{eqnarray}
that defines $f(\widetilde{\xi })$ and $\widetilde{l}.$ They can be found
numerically by discretization as described above. The expression for $%
n^{(0)}(\widetilde{\xi })$ in terms of $f(\widetilde{\xi })$ following from
Eq.\ (\ref{fuDef}) has the form
\begin{equation}
n^{(0)}(\widetilde{\xi })=\frac{1}{2}\left( 1+\frac{\widetilde{W}_{\mathrm{%
ext}}}{\bar{D}_{zz}^{(\mathrm{cyl})}}\right) +\frac{1}{2}\left( 1-\frac{%
\widetilde{W}_{\mathrm{ext}}}{\bar{D}_{zz}^{(\mathrm{cyl})}}\right) f(%
\widetilde{\xi }).  \label{n0Res}
\end{equation}
Using the method of Ref.\ \onlinecite{garchu08prb}, one can show that the
approximate solution for $f(\widetilde{\xi })$ valid for $\left| \widetilde{%
\xi }\right| \ll 1$ has the form
\begin{equation}
f_{\mathrm{app}}(\widetilde{\xi })=\widetilde{\xi }/\widetilde{l}_{\mathrm{%
app}},  \label{fAppr}
\end{equation}
where
\begin{equation}
\widetilde{l}_{\mathrm{app}}=\frac{k}{\sqrt{\left( 4\pi \nu \right)
^{2}-k^{2}}}.  \label{lAppr}
\end{equation}
For a Mn$_{12}$ Ac cylinder one obtains $\widetilde{l}_{\mathrm{app}}=0.7137.
$
\begin{figure}[tbp]
\includegraphics[angle=-90,width=8cm]{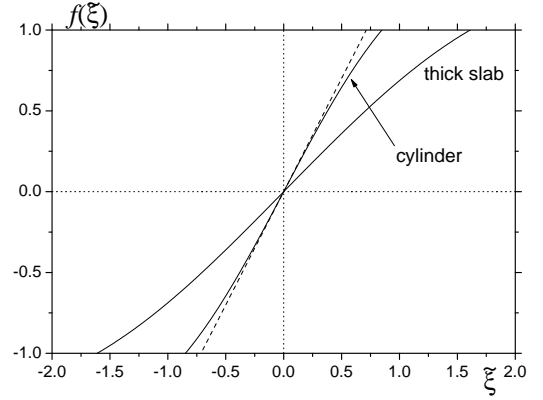}
\caption{Normalized magnetization profile in the cold deflagration front for
long Mn$_{12}$ Ac crystals of cylindrical and thick-slab shape. Approximate
result for the cylinder is shown by a dashed line}
\label{Fig-fxitilde}
\end{figure}
\begin{figure}[tbp]
\includegraphics[angle=-90,width=8cm]{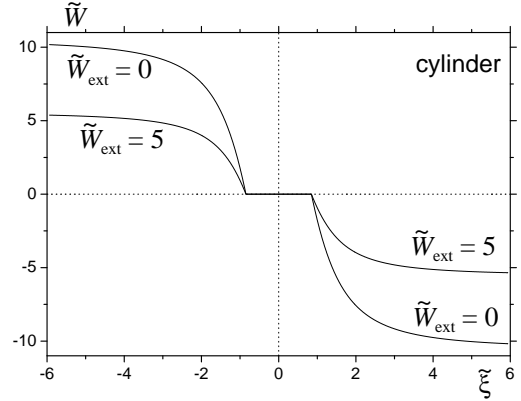}
\caption{Normalized energy bias $\widetilde{W}$ in the cold deflagration
front in long Mn$_{12}$ Ac crystals in the limit of strong dipolar field. In
the central part of the front $\widetilde{W}$ $\cong 0$ so that resonance
transitions take place.}
\label{Fig-Wxitilde}
\end{figure}

Numerically found $f(\widetilde{\xi })$ \ and its approximation $f_{\mathrm{%
app}}(\widetilde{\xi })$ for a Mn$_{12}$ Ac cylinder is shown in Fig.\ \ref
{Fig-fxitilde} together with $f(\widetilde{\xi })$ for a thick slab (ribbon)
discussed at the end of this section.

The speed of the front $v^{\ast }$ can be found by considering the effect of
the correction $\sigma _{z}^{(1)}(\widetilde{\xi })$, although one does not
need to evaluate this correction explicitly. In Eq.\ (\ref{dndzEq}) one has
to keep $\sigma _{z}^{(1)}(\widetilde{\xi })$ in $F,$ because otherwise the
denominator turns to zero. This makes $F$ zero order in $1/\widetilde{E}_{D}.
$ On the other hand, $dn/d\widetilde{\xi }$ and $n$ outside $F$ can be taken
at zero order in $1/\widetilde{E}_{D}$. One thus can rewrite Eq.\ (\ref
{dndzEq}) in the interval $-\widetilde{l}<\widetilde{\xi }<\widetilde{l}$ in
the form
\begin{equation}
\frac{v^{\ast }}{F}=g(\widetilde{\xi }),\qquad g(\widetilde{\xi })\equiv
n^{(0)}\left( \frac{dn^{(0)}}{d\widetilde{\xi }}\right) ^{-1}.
\label{gxitilDef}
\end{equation}
There is a point inside the interval $-\widetilde{l}<\widetilde{\xi }<%
\widetilde{l}$ where $W$ changes its sign. At this point $1/F$ reaches its
minimal value 1. On the other hand, this point can be determined as the
minimum of the rhs of this equation. Then, obviously,
\begin{equation}
v^{\ast }=\min \left[ g(\widetilde{\xi })\right] .  \label{vStarGeneral}
\end{equation}
Using Eq.\ (\ref{n0Res}) one obtains
\begin{equation}
g(\widetilde{\xi })=\frac{Q+1+f(\widetilde{\xi })}{f^{\prime }(\widetilde{%
\xi })},\qquad Q\equiv \frac{2\widetilde{W}_{\mathrm{ext}}}{\bar{D}_{zz}^{(%
\mathrm{cyl})}-\widetilde{W}_{\mathrm{ext}}}.  \label{QDef}
\end{equation}
One can see that for $\widetilde{W}_{\mathrm{ext}}=0$ one has $\min \left[ g(%
\widetilde{\xi })\right] =0,$ achieved at $\widetilde{\xi }=\widetilde{l}$
where $f=-1.$ This yields $v^{\ast }=0$ at $\widetilde{W}_{\mathrm{ext}}=0.$
In the general case one has to investigate
\begin{equation}
g^{\prime }(\widetilde{\xi })=1-\frac{f^{\prime \prime }(\widetilde{\xi })}{%
f^{\prime 2}(\widetilde{\xi })}\left[ Q+1+f(\widetilde{\xi })\right] .
\label{gder}
\end{equation}
Since $g^{\prime }(-\widetilde{l})=0$ at
\begin{equation}
Q=Q_{c}=f^{\prime 2}(-\widetilde{l})/f^{\prime \prime }(-\widetilde{l}),
\label{QcDef}
\end{equation}
one concludes that for $Q\leq Q_{c}$ the minimum is achieved at $\widetilde{%
\xi }=-\widetilde{l}$ and thus
\begin{equation}
v^{\ast }=\frac{Q}{f^{\prime }(-\widetilde{l})}=\frac{\widetilde{W}_{\mathrm{%
ext}}}{\bar{D}_{zz}^{(\mathrm{cyl})}-\widetilde{W}_{\mathrm{ext}}}\frac{2}{%
f^{\prime }(-\widetilde{l})}.  \label{vStarRes}
\end{equation}
For a cylinder of Mn$_{12}$ Ac one has $2/f^{\prime }(-\widetilde{l})=2.31$
and $Q_{c}=0.809.$ According to Eq.\ (\ref{QDef}), the latter translates
into $\widetilde{W}_{\mathrm{ext,}c}=3.03.$ Then Eq.\ (\ref{WtilExtFinal})
yields the value of the corresponding bias field $B_{z,c}-B_{zk}=15$ mT. For
$B_{zk}\leq B_{z}\leq B_{z,c},$ the front speed in real units obtained with
the help of Eqs.\ (\ref{WtilExtFinal}), (\ref{BDiz}), and (\ref{EDDef}) is
given by
\begin{equation}
v=R\Gamma _{\mathrm{res}}\frac{B_{z}-B_{zk}}{B_{z}^{(D)}-B_{z}+B_{zk}}\frac{2%
}{f^{\prime }(-\widetilde{l})}.  \label{vstarRealUnits}
\end{equation}
For $Q_{c}\leq Q$ (and thus $\widetilde{W}_{\mathrm{ext,}c}\leq \widetilde{W}%
_{\mathrm{ext}})$ one has to find $\min \left[ g(\widetilde{\xi })\right] $
from the condition $g^{\prime }(\widetilde{\xi })=0$ that leads to somewhat
smaller front speeds than given by the formulas above. However, the laminar
regime of the cold deflagration fronts breaks down at the external bias
smaller than $\widetilde{W}_{\mathrm{ext,}c},$ so that the results of this
section for $\widetilde{W}_{\mathrm{ext,}c}\leq \widetilde{W}_{\mathrm{ext}}$
are irrelevant.

Let now consider the slab geometry. From Eq.\ (\ref{DzzThickSlab}) with $%
\widetilde{z}\equiv 2z/a$ instead of Eq.\ (\ref{ZeroApproxW}) one obtains
the equation
\begin{eqnarray}
&&\widetilde{W}^{(0)}(\widetilde{\xi })=\widetilde{W}_{\mathrm{ext}}+\int_{-%
\widetilde{l}}^{\widetilde{l}}d\widetilde{\xi }^{\prime }\frac{4\nu \sigma
_{z}^{(0)}(\widetilde{\xi }^{\prime })}{\left( \widetilde{\xi }^{\prime }-%
\widetilde{\xi }\right) ^{2}+1}  \nonumber \\
&&{}-k\sigma _{z}^{(0)}(\widetilde{\xi })+4\nu \left[ \sigma _{zi}\left(
\frac{\pi }{2}+\mathrm{Arctan}\left( \widetilde{\xi }-\widetilde{l}\right)
\right) \right.  \nonumber \\
&&+\left. \sigma _{zf}\left( \frac{\pi }{2}-\mathrm{Arctan}\left( \widetilde{%
\xi }+\widetilde{l}\right) \right) \right] .  \label{ZeroApproxWSlab}
\end{eqnarray}
The numerically obtained result for the normalized magnetization profile $f(%
\widetilde{\xi })$ is shown of Fig.\ \ref{Fig-fxitilde}, compared to that of
a cylinder. Since the kernel in the integral equation for the slab is less
localized for a thick slab than for a cylinder, the front width $\widetilde{l%
}=1.61$ for a thick slab is larger than $\widetilde{l}=0.848$ for the
cylinder. All formulas obtained above are valid for a thick slab as well,
however with different constants: $2/f^{\prime }(-\widetilde{l})=4.79$, $%
Q_{c}=0.540,$ and $\widetilde{W}_{\mathrm{ext,}c}=2.24.$

\section{Numerical results}

\label{Sec-numerical}

\begin{figure}[tbp]
\includegraphics[angle=-90,width=8cm]{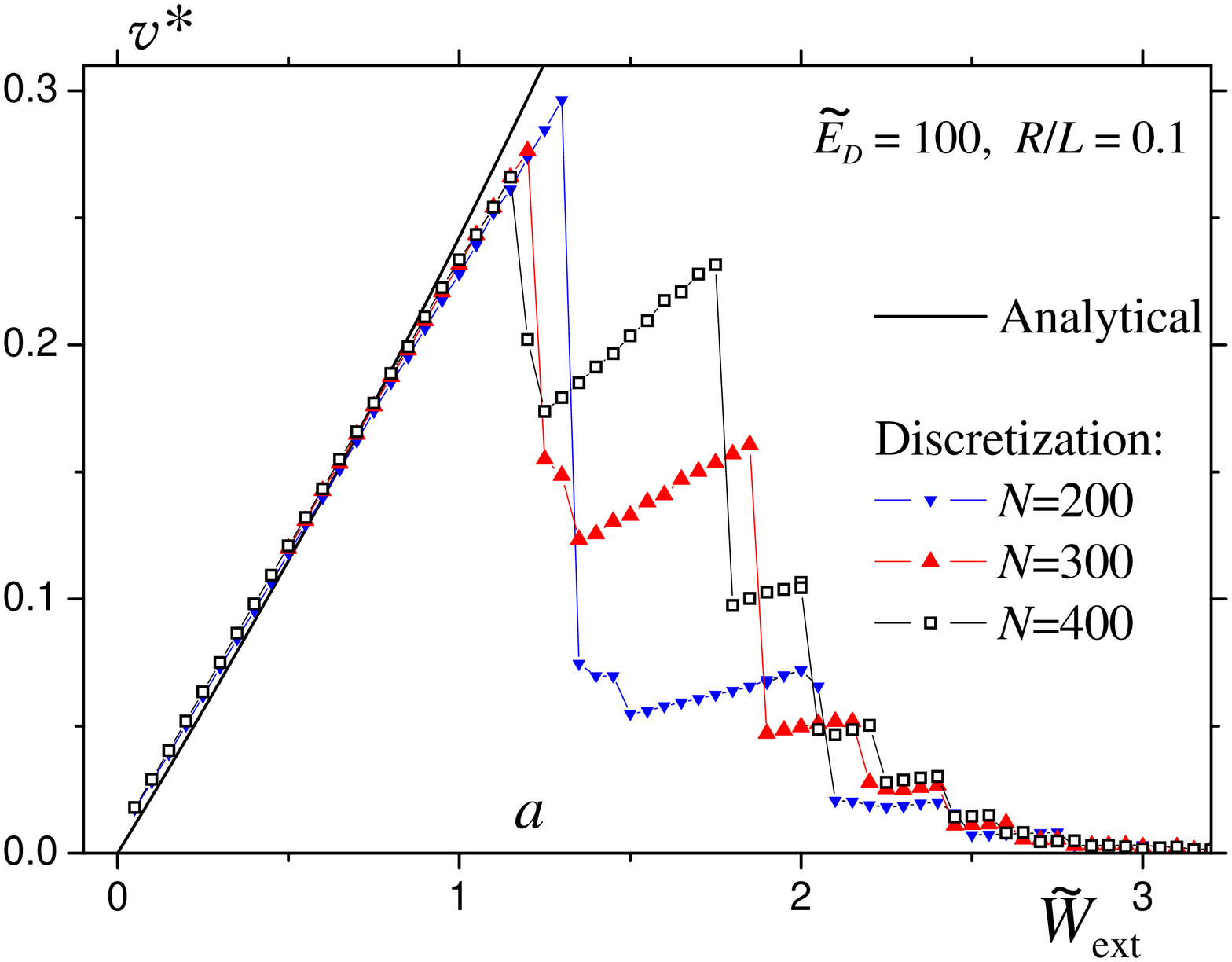} %
\includegraphics[angle=-90,width=8cm]{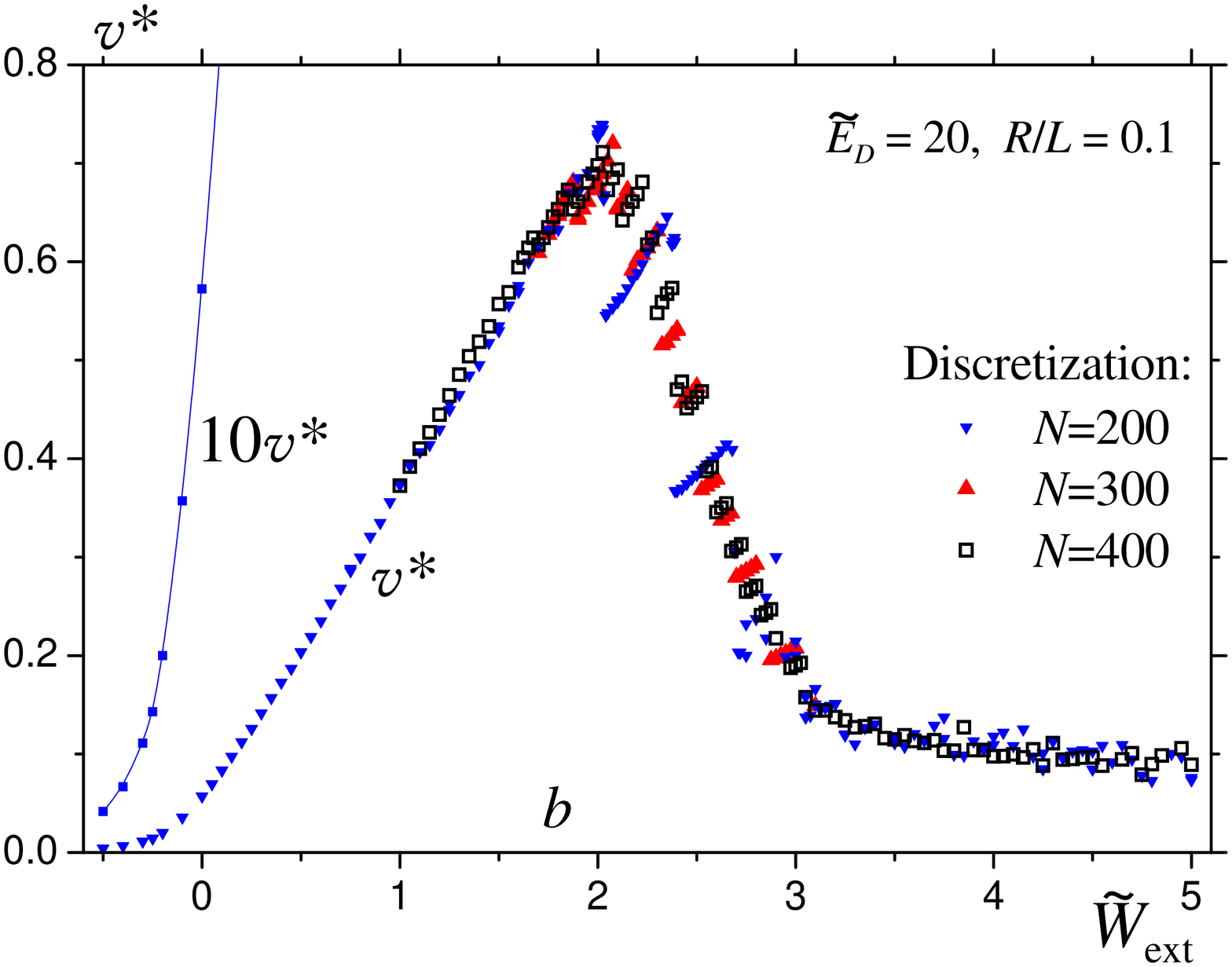}
\caption{Reduced front speed $v^{\ast }$ vs the reduced bias $\widetilde{W}_{%
\mathrm{ext}}$ for different discretizations. In (a) for $\widetilde{E}%
_{D}=100$ and $\widetilde{W}_{\mathrm{ext}}\lesssim 1$ the numerical results
are in a good accordance with the large-$\widetilde{E}_{D}$ formula
(straight line).}
\label{Fig_vstar_vs_bias}
\end{figure}

As mentioned at the end of Sec. \ref{Sec-cold-deflagr-eqs}, the
cold-deflagration equations can be solved by discretization reducing them to
a system of ordinary differential equation. Numerical calculations use the
semi-infinite geometry including the region of length $-L/2\leq z\leq L/2$
where equations are solved plus the range $L/2<z\leq \infty $ where the
magnetization is fixed to $\sigma _{z}=-1$ corresponding to the metastable
state. The latter is needed to create the dipolar field in the main region $%
-L/2\leq z\leq L/2$ that corresponds to the semi-infinite sample. This
allows to operate on shorter samples that saves computer time. When the
deflagration front reaches $z=L/2,$ it cannot go further, so the results
near this point become unphysical and should be ignored.

The first thing revealed by computations is that for large values of $%
\widetilde{E}_{D}$ it is very important to prepare the system in the initial
state close to the actual front, with $\widetilde{W}\approx 0$ within the
front core. The further is the initial state from this optimal state, the
more time (ignition time) the system needs to adjust so the the front could
start moving across the sample. For initial states far from the front
states, the ignition time can be so long that there is a significant
off-resonance relaxation in the bulk of the crystal during it. For smaller
dipolar fields such as $\widetilde{E}_{D}\sim 3$ (that can be achieved by
applying a strong transverse field to increase $\Delta $) ignition of the
fronts is much easier. Another possibility to ignite the cold deflagration
is to slowly sweep the external field in the positive direction, approaching
the resonance, \cite{garchu09prl} that will be considered later on.

Computations for large $\widetilde{E}_{D}$ and not too strong bias $%
\widetilde{W}_{\mathrm{ext}}$ corroborate semi-analytical results of the
preceding section. For $z$ not too close to the ends of the interval $%
-L/2\leq z\leq L/2,$ the variables indeed depend on $\xi =z-vt,$ as it
should be in a moving front. 3D plots of $n(z,t)$ are smooth and look
qualitatively as in Fig.\ \ref{Fig-n_3d_ED=20_Wext=2}, and the ignition time
can be reduced to zero by a better choice if the initial state. The
metastable population $\bar{n}(t)$ averaged over the length of the crystal
is almost flat during the ignition time, then it decreases linearly as the
front travels through the crystal, then becomes nearly flat again after the
front arrives at the right end of the interval $-L/2\leq z\leq L/2$, see
Fig.\ 16 of Ref.\ \onlinecite{garchu07prb} for the standard magnetic
deflagration and Fig.\ 4 of Ref.\ \onlinecite{garchu09prl} for the cold
deflagration. The front speed can be obtained as $v=L/\left( t_{\mathrm{%
arrival}}-t_{\mathrm{ignition}}\right) .$

The reduced front speed $v^{\ast },$ Eq.\ (\ref{vviavstar}), vs $\widetilde{W%
}_{\mathrm{ext}},$ Eq.\ (\ref{WtilExtFinal}), is shown for $\widetilde{E}%
_{D}=100$ in Fig.\ \ref{Fig_vstar_vs_bias}a. For not too large bias, $%
\widetilde{W}_{\mathrm{ext}}\lesssim 1,$ the numerical results are in a good
agreement with Eq.\ (\ref{vStarRes}) shown by a solid line. In this region
the numerical results do not depend on the number of discrete points used in
the solution. Similar results for $\widetilde{E}_{D}=20$ in Fig.\ \ref
{Fig_vstar_vs_bias}b are further from the theoretical curve (not shown)
because the condition of a strong dipolar field is not fully satisfied. Also
there is some nonzero speed in the region $\widetilde{W}_{\mathrm{ext}}<0$
that is, however, quickly decreasing with the negative bias.

\begin{figure}[tbp]
\includegraphics[angle=-90,width=8cm]{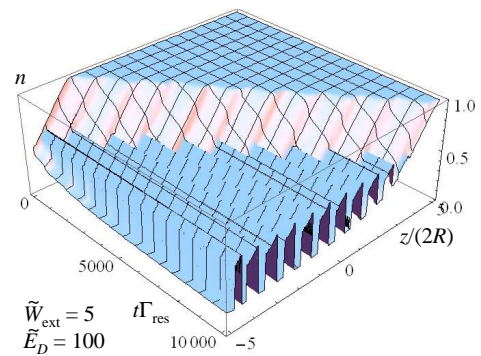}
\caption{{}Propagating front of cold deflagration for $\widetilde{E}_{D}=100$
and $\widetilde{W}_{\mathrm{ext}}=5$. Here ignition time was eliminated by
good choice of the initial condition. The front speed is oscillating and
there are spatially-periodic structures behind the front.}
\label{Fig-n_3d_ED=100_Wext=5}
\end{figure}
\begin{figure}[tbp]
\includegraphics[angle=-90,width=8cm]{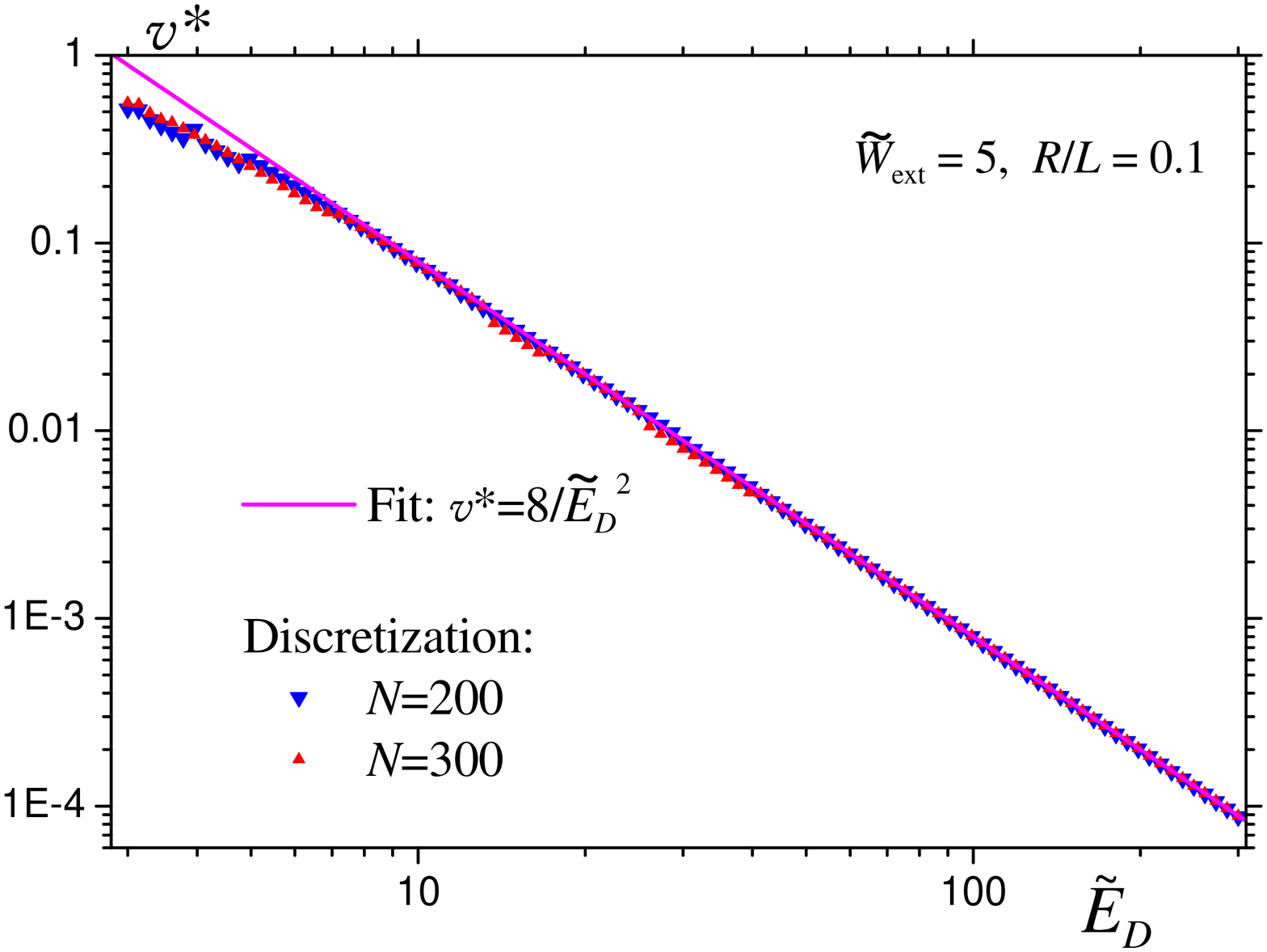}
\caption{{}Front speed $v^{\ast }$ vs $\widetilde{E}_{D}$ in the non-laminar
regime for $\widetilde{W}_{\mathrm{ext}}=5.$}
\label{Fig_vstar_vs_ED}
\end{figure}

With increasing the external bias, the laminar solution in the form of a
smooth moving front loses its stability. In Fig.\ \ref
{Fig-n_W_profiles_ED=20_Wext=2} one can already see wiggles behind the front
that represent frozen-in spatial structures with the period of order $R$.
With increasing $\widetilde{W}_{\mathrm{ext}}$ or $\widetilde{E}_{D},$ these
features progress and the region of transitions moves with oscillating
speed, see Fig.\ \ref{Fig-n_3d_ED=100_Wext=5}. To distinguish this
transition region from the true front, it was called ``wall of
transitions''. \cite{garchu09prl} It would cost significant additional
efforts to find out analytically or numerically whether the transition from
the laminar to non-laminar regime with increasing $\widetilde{W}_{\mathrm{ext%
}}$ is gradual or there is a threshold. One important observation is that
the spatial structures behind the front are discontinuous on $z,$ while the
analytical result of Eq.\ (\ref{vStarGeneral}) was obtained based on the
assumption that the solution is continuous.

As the laminar regime breaks down, the instability is manifested by the
dependence of the results for $v^{\ast }$ in Fig.\ \ref{Fig_vstar_vs_bias}
on the number of discretization points. With increasing $N$ the
discontinuities in $v^{\ast }(\widetilde{W}_{\mathrm{ext}})$ become smaller
but, unfortunately, increasing $N$ is limited by computing resourses. For
larger $\widetilde{W}_{\mathrm{ext}}$ the front speed reaches a plateau that
depends on $\widetilde{E}_{D},$ and curves with different discretizations
converge again.

In the non-laminar regime the magnetization in the front cannot completely
adjust so that bias in the front core would be very close to zero and the
resonance transitions could occur at a rate close to the maximal rate $%
\Gamma _{\mathrm{res}}.$ This is the reason why $v^{\ast }(\widetilde{W}_{%
\mathrm{ext}})$ drops after reaching a maximum, as the instability begins.
Still the very existence of the front in this case suggests that the system
is closer to the resonance in this region than in the others. The values of $%
\widetilde{W}$ in the front should be of order 1, so that in Eq.\ (\ref{FDef}%
) one has $F\sim 1/\widetilde{E}_{D}^{2}.$ This is supported by the
computations shown in Fig.\ \ref{Fig_vstar_vs_ED}: In the plateau region (in
particular for $\widetilde{W}_{\mathrm{ext}}=5)$ the front speed fits to
\begin{equation}
v^{\ast }\simeq 8/\widetilde{E}_{D}^{2}  \label{vstarvsED}
\end{equation}
for large $\widetilde{E}_{D}.$

\section{Cold deflagration initiated by field sweep}

\label{Sec-Sweep}

\begin{figure}[tbp]
\includegraphics[angle=-90,width=8cm]{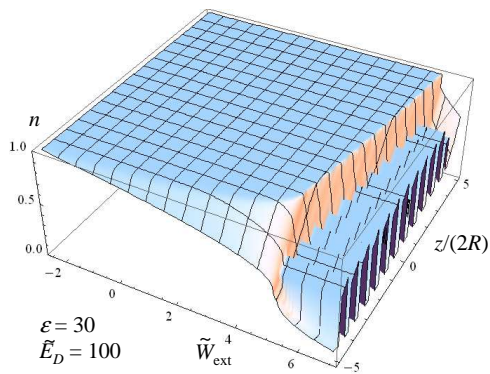}
\caption{{}Wall of tunneling at $\widetilde{E}_{D}=100,$ ignited by slow
sweep of the bias field, $\protect\varepsilon =30.$ The process starts at $%
\widetilde{W}_{\mathrm{ext}}\approx 5.$}
\label{Fig-n_3d_ED=100_Sweep}
\end{figure}
\begin{figure}[tbp]
\includegraphics[angle=-90,width=8cm]{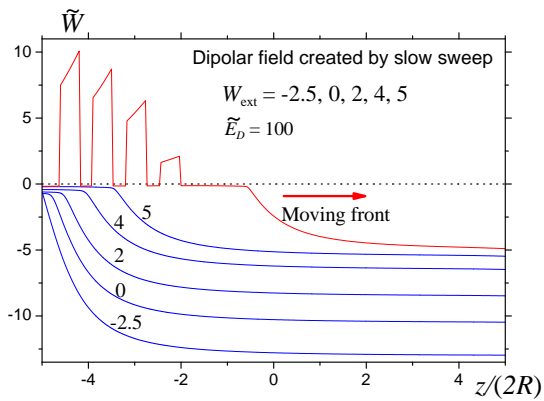}
\caption{{}Dipolar field in the crystal at different values of $\widetilde{W}%
_{\mathrm{ext}}$\ during sweep. After $\widetilde{W}_{\mathrm{ext}}=5$ the
front starts moving and dipolar field in the sample becomes discontinuous.}
\label{Fig-Sweep-profiles}
\end{figure}

As was mentioned above, for most of initial conditions the development of
the cold deflagration front requires a very long ignition time. If the
initial condition is chosen in a special way close to the actual front, the
process starts immediately. However, one cannot find a practical way to
prepare such initial state.

Fortunately, as was found in Ref.\ \onlinecite{garchu09prl}, the front can
be ignited by a slow time-linear sweep $W_{\mathrm{ext}}=v_{W}t$ starting
with a value of $W_{\mathrm{ext}}$ that ensures $W<0$ everywhere in the
sample, see Fig.\ \ref{Fig-n_3d_ED=100_Sweep}. The sweep rate can be
conveniently parametrized by $\varepsilon \equiv \pi \Delta ^{2}/\left(
2\hbar v_{W}\right) $ and slow sweep requires $\varepsilon \gg 1.$ As $W_{%
\mathrm{ext}}$ increases, the condition $W=0$ would be first reached at the
end of the sample, then the resonance point would move into its depth.
However, transitions induced by the sweep (seen in Fig.\ \ref
{Fig-n_3d_ED=100_Sweep} before the ignition) change the dipolar field so
that the system does not cross the resonance in the region near the end of
the sample, although it becomes close to the resonance in the increasingly
broad region, see Fig.\ \ref{Fig-Sweep-profiles}. The reason for this is
that flipping spins in a small region near the end of the sample do not
significantly change the dipolar field from the surface of the crystal, the
integral term of Eq.\ (\ref{DzzCylinder}), but strongly change the local
contribution, last term in this formula. Increasing $\sigma _{z}$ due to
resonance transitions leads to the \emph{decrease} of the local term that
creates a negative dipolar bias and prevents the system from crossing the
resonance. After some time the region close to the resonance becomes broad
that is similar to the structure of the cold deflagration front, see Fig.\
\ref{Fig-Sweep-profiles}. In this way the initial state for the cold
deflagration is being prepared. The front starts as the bias reaches the
``magic'' value of $\widetilde{W}_{\mathrm{ext}}$ that weakly depends on $%
\widetilde{E}_{D}.$ For $\widetilde{E}_{D}=20$ one has $\widetilde{W}_{%
\mathrm{ext}}=4.3$ (that corresponds to $B_{z}-B_{zk}\simeq 19$ mT) and for $%
\widetilde{E}_{D}=100$ one has $\widetilde{W}_{\mathrm{ext}}\simeq 5.$ At
such a strong bias there is a quasiperiodic spatial structure with
discontinuous magnetization and the dipolar field behind the front, as shown
in Figs.\ \ref{Fig-n_3d_ED=100_Sweep} and \ref{Fig-Sweep-profiles}. One can
see in Fig.\ \ref{Fig-Sweep-profiles} that in the moving front the bias is
slightly below zero. This means that the system is somewhat off-resonance
and this is the reason for a small front speed in this regime, as shown in
Fig.\ \ref{Fig_vstar_vs_bias}.

\begin{figure}[tbp]
\includegraphics[angle=-90,width=8cm]{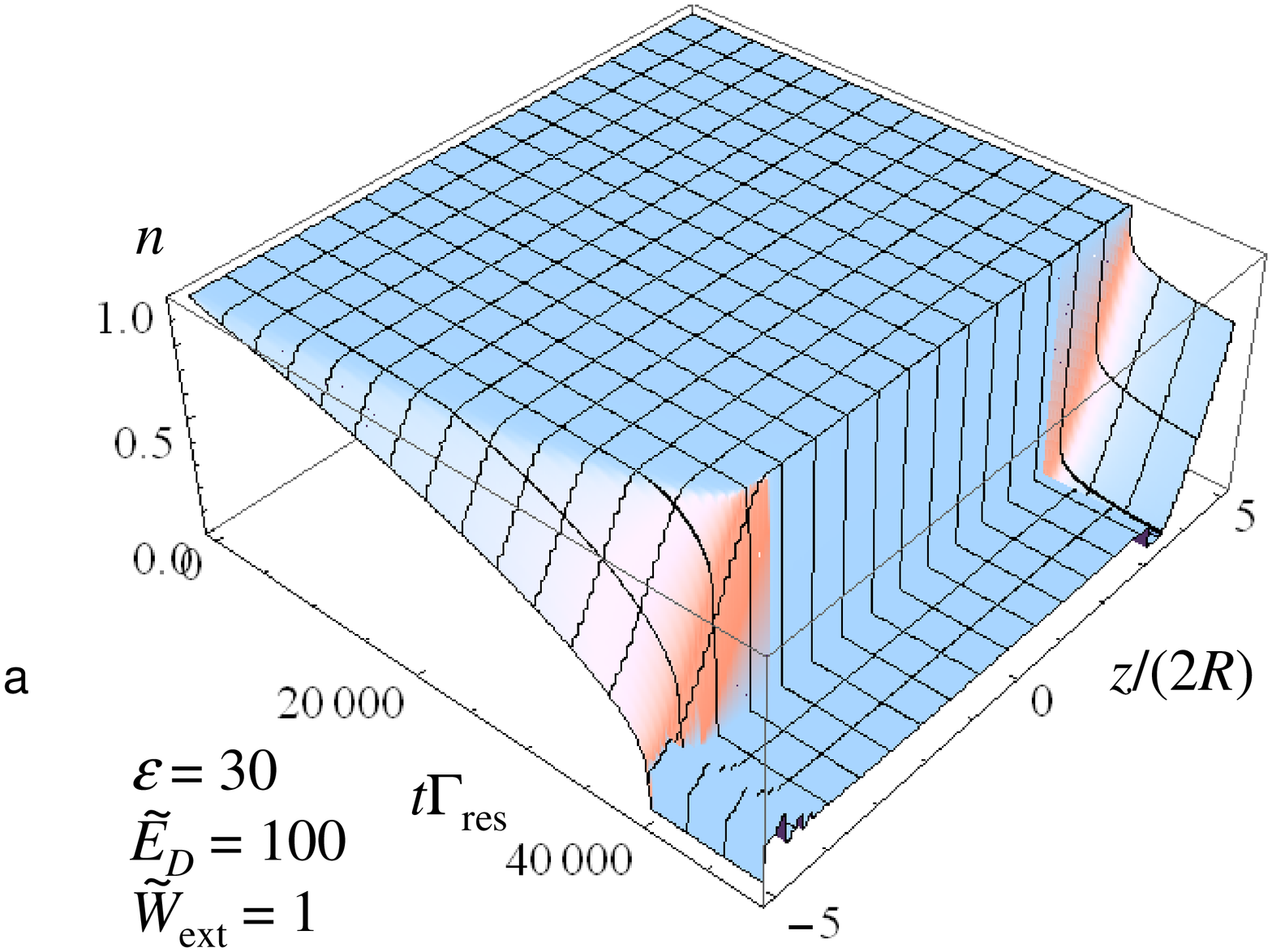} %
\includegraphics[angle=-90,width=8cm]{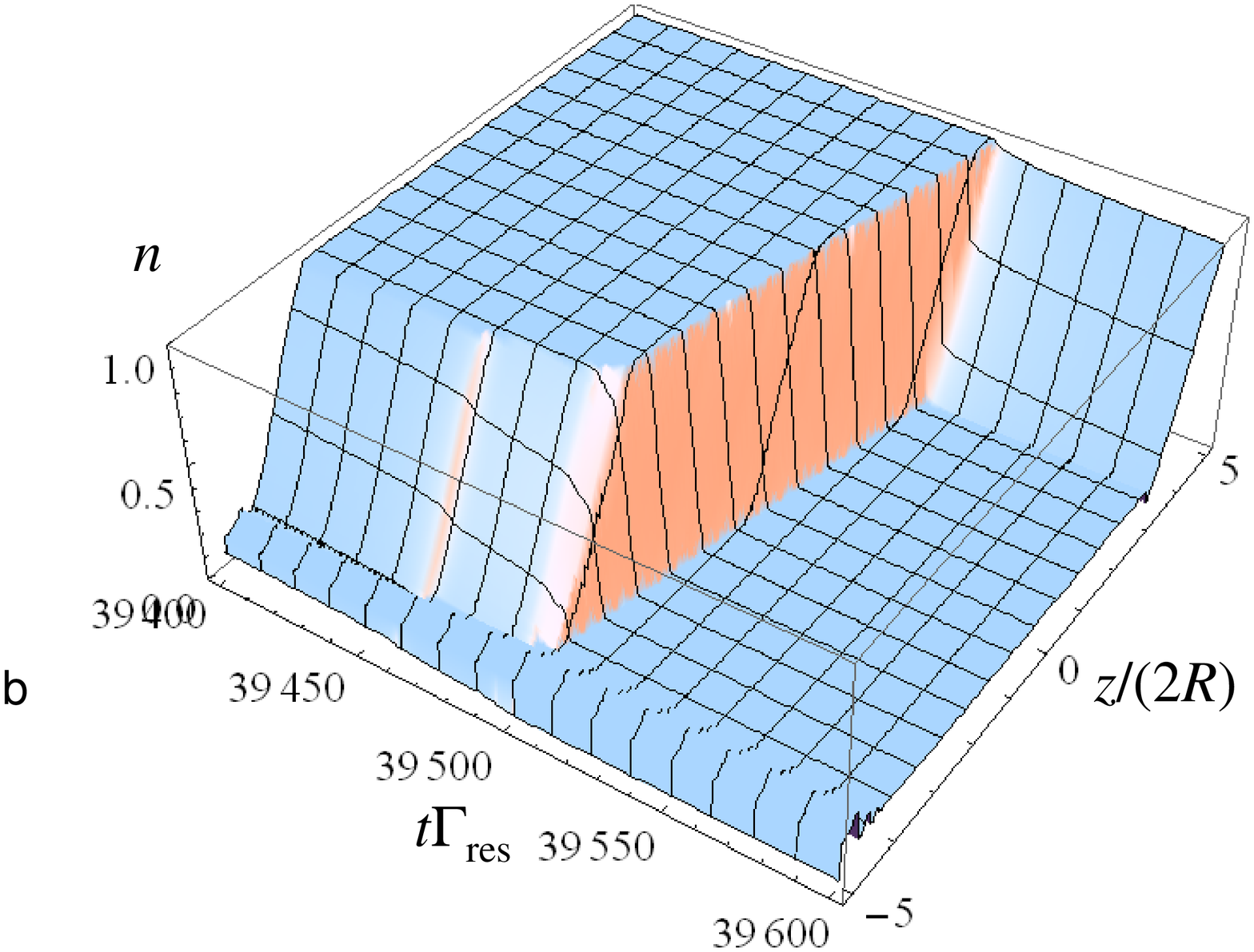}
\caption{{}Laminar front of tunneling at $\widetilde{E}_{D}=100$ and $%
\widetilde{W}_{\mathrm{ext}}=1$ ignited by slow \emph{local} sweep of the
bias field. (a) Overview; (b) Zoom of the front region.}
\label{Fig-n_3d_ED=100_Local_Sweep}
\end{figure}
\begin{figure}[tbp]
\includegraphics[angle=-90,width=8cm]{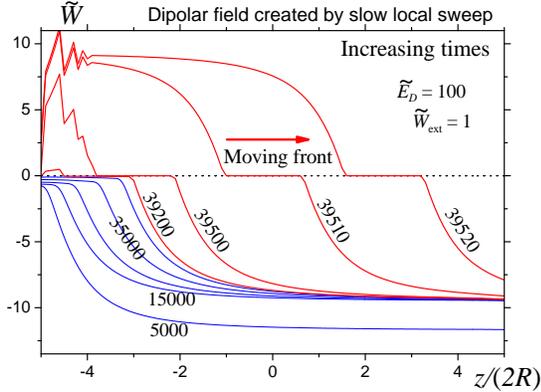}
\caption{{}Total bias at different times in the laminar front of tunneling
at $\widetilde{E}_{D}=100$ and $\widetilde{W}_{\mathrm{ext}}=1$ ignited by
slow \emph{local} sweep of the bias field. Blue lines correspond to times of
adjustment before ignition and red lines correspond to the moving front.}
\label{Fig-n_ED=100_Local_Sweep_profiles}
\end{figure}

The next question is how to ignite cold deflagration for arbitrary values of
the external bias $\widetilde{W}_{\mathrm{ext}}$ given by Eq.\ (\ref
{WtilExtFinal}). The answer is to sweep $B_{z}$ up to this value of $%
\widetilde{W}_{\mathrm{ext}}$ and then to stop this (global) sweep at some $%
t_{\max ,0}$. After that further sweep $B_{z}$ locally near the end of the
sample, $z=-L/2,$ using a small coil. For the coil of radius $R_{0}$ placed
at $z_{0}$ (the axis of the coil coincides with the axis of the cylinder)
the local addition to $W_{\mathrm{ext}}$ can be written in the form
\begin{equation}
\delta W_{\mathrm{ext}}(z,t)=\frac{R_{0}^{2}}{R_{0}^{2}+\left(
z-z_{0}\right) ^{2}}v_{W}\left( t-t_{\max ,0}\right)   \label{deltaWext}
\end{equation}
with $z_{0}\approx -L/2.$ Numerical calculations with $R_{0}=R$ and $%
z_{0}=-L/2$ show that, indeed, with this method one can ignite fronts at
different biases $\widetilde{W}_{\mathrm{ext}},$ including the laminar
regime, see Fig.\ \ref{Fig-n_3d_ED=100_Local_Sweep}. Here the front is much
faster than in Fig.\ \ref{Fig-n_3d_ED=100_Sweep}, in accordance with the
results for the front speed in Fig.\ \ref{Fig_vstar_vs_bias}a. Total bias in
the sample at different times is shown in Fig.\ \ref
{Fig-n_ED=100_Local_Sweep_profiles}. For instance, $\tau \equiv \Gamma _{%
\mathrm{res}}t=5000$ corresponds to the stage of the global sweep. All other
times correspond to the local-sweep stage, since the bias curves converge on
the right side of the sample where $\delta W_{\mathrm{ext}}$ is small. Local
sweep near the left end creates an initial state for the front to start, as
the bias curves are approaching zero in a progressively large region (blue
curves). As the front starts moving (red curves), the bias becomes positive
on the left with non-laminar features near the end. But in the depth of the
sample the front is laminar corresponding to $\widetilde{W}_{\mathrm{ext}}=1.
$

\section{Discussion}

\label{Sec-Discussion}

In the main part of the paper it was shown that elongated crystals of
molecular magnets (practically Mn$_{12}$ Ac) can exhibit moving fronts of
dissipative spin tunneling at biased resonances, Eq.\ (\ref{BkDef}) with $%
k>0.$ This phenomenon is resembling magnetic deflagration,\cite
{suzetal05prl,garchu07prb} only the relaxation rate is controlled not by the
temperature but by the dipolar field evolving self-consistently and bringing
the spins in the front region on and off tunneling resonance. Like
deflagration, it leads to destruction of the initial metastable ordered
state (however in general incomplete), this is why it can be called ``cold
deflagration''.

Of course, transitions at biased resonances result in energy release and
warming of the sample, so that the two mechanisms can coexist. In fact,
magnetic deflagration was observed in crystals of Mn$_{12}$ Ac thermally
isolated so that the warming of the sample is efficient. Without thermal
isolation, cold deflagration does not face this competition. To further
reduce heating, it is preferable to work at low bias, such as $k=1.$

There are two regimes of cold deflagration: laminar regime at low bias $%
B_{z}-B_{zk}$ and non-laminar regime at high bias. In the laminar regime the
magnetization in the front adjusts so that the dipolar field $B^{(D)}$
together with the external field $B_{z}$ creates a nearly zero bias for the
resonance transitions in the front region with the width of order $R,$ the
transverse size of the crystal. In the laminar regime the magnetization and
dipolar field in the sample are continuous and both the front speed and the
magnetization (metastable population) behind the front can be found
analytically in the practical limit of the strong dipolar field, Eqs.\ (\ref
{vstarRealUnits}) and (\ref{nFinal}). Remarcably, both of these quantities
do not depend on the strength of the dipolar field $E_{D}$ in this region.

In the laminar regime the estimation for the front speed is $v\sim R\Gamma
_{res},$ where $\Gamma _{res}$ is the transition rate at resonance, $W=0$ in
Eq.\ (\ref{GammaDef}). At the boundary between the over- and underdamped
regimes, $\Delta \sim \Gamma _{m^{\prime }}$ and thus $\Gamma _{res}\sim
\Gamma _{m^{\prime }}$ (that is realized in the transverse field $3$ T in Mn$%
_{12}$ Ac at the $k=1$ resonance) cold deflagration already beats the
regular ``hot'' deflagration. Indeed, the latter has the speed $v\sim
l\Gamma (T_{f}),$ where $l$ depends on the thermal diffusivity but
experimentally is comparable with $R$ and $\Gamma (T_{f})$ is the thermal
activation rate over the barrier at the flame temperature. Since $\Gamma (T)$
at high temperatures is determined by the rates of transitions between
adjacent levels near the top of the barrier that are smaller than the same
for low-lying levels such as $\Gamma _{m^{\prime }}$ (one has $\Gamma
(T_{f})\sim 10^{6}$ s$^{-1}$ and $\Gamma _{S-1}\sim 10^{7}$ s$^{-1})$ the
hot deflagration loses.

At higher bias $B_{z}-B_{zk}$ the laminar regime breaks down, the dipolar
field cannot fully adjust to provide a nearly zero bias in the front's core,
and the magnetization and the dipolar field become discontinuous. There are
frozen-in quasiperiodic spatial structures behind the front. Accordingly,
the front speed dramatically drops, see Fig.\ \ref{Fig_vstar_vs_bias},
especially in the case of a strong dipolar field. There is no analytical
solution in this range but the fit to the numerical results yields $%
v\varpropto 1/E_{D}^{2}.$ The boundary between the laminar and non-laminar
regimes corresponds to $B_{z}-B_{zk}=5$-10 mT.

It was shown that cold deflagration can be ignited by the local sweep of the
field $B_{z}$ near an end of the crystal. This local field can be produced
by a small coil with increasing current.

Another condition for the observability of the cold deflagration is
sufficiently strong transverse field that allows tunneling transitions via
low-lying levels.

At nonzero temperatures the rate of cold deflagration should increase
because of the activation to higher levels providing a higher transition
probability, see Eq.\ (10) of Ref.\ \onlinecite{garchu09prl}.

\section*{Acknowledgements}

The author is indebted to E. M. Chudnovsky for many stimulating discussions.

This work has been supported by the NSF Grant No. DMR-0703639.

\bibliographystyle{apsrev}
\bibliography{gar-own,chu-own,gar-tunneling,gar-relaxation,gar-oldworks,gar-books,gar-general}

\begin{thebibliography}{24}
\expandafter\ifx\csname natexlab\endcsname\relax\def\natexlab#1{#1}\fi
\expandafter\ifx\csname bibnamefont\endcsname\relax
  \def\bibnamefont#1{#1}\fi
\expandafter\ifx\csname bibfnamefont\endcsname\relax
  \def\bibfnamefont#1{#1}\fi
\expandafter\ifx\csname citenamefont\endcsname\relax
  \def\citenamefont#1{#1}\fi
\expandafter\ifx\csname url\endcsname\relax
  \def\url#1{\texttt{#1}}\fi
\expandafter\ifx\csname urlprefix\endcsname\relax\def\urlprefix{URL }\fi
\providecommand{\bibinfo}[2]{#2}
\providecommand{\eprint}[2][]{\url{#2}}

\bibitem[{\citenamefont{Lis}(1980)}]{lis80}
\bibinfo{author}{\bibfnamefont{T.}~\bibnamefont{Lis}}, \bibinfo{journal}{Acta
  Crystallogr. B} \textbf{\bibinfo{volume}{36}}, \bibinfo{pages}{2042}
  (\bibinfo{year}{1980}).

\bibitem[{\citenamefont{R.~Sessoli and Novak}(1993)}]{sesgatcannov93nat}
\bibinfo{author}{\bibfnamefont{A.~C.} \bibnamefont{R.~Sessoli},
  \bibfnamefont{D.~Gatteschi}} \bibnamefont{and}
  \bibinfo{author}{\bibfnamefont{M.~A.} \bibnamefont{Novak}},
  \bibinfo{journal}{Nature (London)} \textbf{\bibinfo{volume}{365}},
  \bibinfo{pages}{141} (\bibinfo{year}{1993}).

\bibitem[{\citenamefont{Friedman et~al.}(1996)\citenamefont{Friedman, Sarachik,
  Tejada, and Ziolo}}]{frisartejzio96prl}
\bibinfo{author}{\bibfnamefont{J.~R.} \bibnamefont{Friedman}},
  \bibinfo{author}{\bibfnamefont{M.~P.} \bibnamefont{Sarachik}},
  \bibinfo{author}{\bibfnamefont{J.}~\bibnamefont{Tejada}}, \bibnamefont{and}
  \bibinfo{author}{\bibfnamefont{R.}~\bibnamefont{Ziolo}},
  \bibinfo{journal}{Phys. Rev. Lett.} \textbf{\bibinfo{volume}{76}},
  \bibinfo{pages}{3830} (\bibinfo{year}{1996}).

\bibitem[{\citenamefont{Hern\'andez et~al.}(1996)\citenamefont{Hern\'andez,
  Zhang, Luis, Bartolom\'e, Tejada, and Ziolo}}]{heretal96epl}
\bibinfo{author}{\bibfnamefont{J.~M.} \bibnamefont{Hern\'andez}},
  \bibinfo{author}{\bibfnamefont{X.~X.} \bibnamefont{Zhang}},
  \bibinfo{author}{\bibfnamefont{F.}~\bibnamefont{Luis}},
  \bibinfo{author}{\bibfnamefont{J.}~\bibnamefont{Bartolom\'e}},
  \bibinfo{author}{\bibfnamefont{J.}~\bibnamefont{Tejada}}, \bibnamefont{and}
  \bibinfo{author}{\bibfnamefont{R.}~\bibnamefont{Ziolo}},
  \bibinfo{journal}{Europhys. Lett.} \textbf{\bibinfo{volume}{35}},
  \bibinfo{pages}{301} (\bibinfo{year}{1996}).

\bibitem[{\citenamefont{Thomas et~al.}(1996)\citenamefont{Thomas, Lionti,
  Ballou, Gatteschi, Sessoli, and Barbara}}]{thoetal96nat}
\bibinfo{author}{\bibfnamefont{L.}~\bibnamefont{Thomas}},
  \bibinfo{author}{\bibfnamefont{F.}~\bibnamefont{Lionti}},
  \bibinfo{author}{\bibfnamefont{R.}~\bibnamefont{Ballou}},
  \bibinfo{author}{\bibfnamefont{D.}~\bibnamefont{Gatteschi}},
  \bibinfo{author}{\bibfnamefont{R.}~\bibnamefont{Sessoli}}, \bibnamefont{and}
  \bibinfo{author}{\bibfnamefont{B.}~\bibnamefont{Barbara}},
  \bibinfo{journal}{Nature} \textbf{\bibinfo{volume}{383}},
  \bibinfo{pages}{145} (\bibinfo{year}{1996}).

\bibitem[{\citenamefont{Morello et~al.}(2003)\citenamefont{Morello, Mettes,
  Luis, Fern\'andez, Krzystek, Aromi, Christou, and de~Jongh}}]{moretal03prl}
\bibinfo{author}{\bibfnamefont{A.}~\bibnamefont{Morello}},
  \bibinfo{author}{\bibfnamefont{E.~L.} \bibnamefont{Mettes}},
  \bibinfo{author}{\bibfnamefont{F.}~\bibnamefont{Luis}},
  \bibinfo{author}{\bibfnamefont{J.~F.} \bibnamefont{Fern\'andez}},
  \bibinfo{author}{\bibfnamefont{J.}~\bibnamefont{Krzystek}},
  \bibinfo{author}{\bibfnamefont{G.}~\bibnamefont{Aromi}},
  \bibinfo{author}{\bibfnamefont{G.}~\bibnamefont{Christou}}, \bibnamefont{and}
  \bibinfo{author}{\bibfnamefont{L.~J.} \bibnamefont{de~Jongh}},
  \bibinfo{journal}{Phys. Rev. Lett.} \textbf{\bibinfo{volume}{90}},
  \bibinfo{pages}{017206} (\bibinfo{year}{2003}).

\bibitem[{\citenamefont{Garanin and Chudnovsky}(2008)}]{garchu08prb}
\bibinfo{author}{\bibfnamefont{D.~A.} \bibnamefont{Garanin}} \bibnamefont{and}
  \bibinfo{author}{\bibfnamefont{E.~M.} \bibnamefont{Chudnovsky}},
  \bibinfo{journal}{Phys. Rev. B} \textbf{\bibinfo{volume}{78}},
  \bibinfo{pages}{174425} (\bibinfo{year}{2008}).

\bibitem[{\citenamefont{{S. McHugh, R. Jaafar, M. P. Sarachik, Y. Myasoedov, H.
  Shtrikman, E. Zeldov, R. Bagai, and G. Christou}}(2009)}]{mchughetal09prb}
\bibinfo{author}{\bibnamefont{{S. McHugh, R. Jaafar, M. P. Sarachik, Y.
  Myasoedov, H. Shtrikman, E. Zeldov, R. Bagai, and G. Christou}}},
  \bibinfo{journal}{Phys. Rev. B} \textbf{\bibinfo{volume}{79}},
  \bibinfo{pages}{052404} (\bibinfo{year}{2009}).

\bibitem[{\citenamefont{Prokof'ev and Stamp}(1998)}]{prosta98prl}
\bibinfo{author}{\bibfnamefont{N.~V.} \bibnamefont{Prokof'ev}}
  \bibnamefont{and} \bibinfo{author}{\bibfnamefont{P.~C.~E.}
  \bibnamefont{Stamp}}, \bibinfo{journal}{Phys. Rev. Lett.}
  \textbf{\bibinfo{volume}{80}}, \bibinfo{pages}{5794} (\bibinfo{year}{1998}).

\bibitem[{\citenamefont{{ A. Cuccoli, A. Fort, A. Rettori, E. Adam, and J.
  Villain}}(1999)}]{cucforretadavil99epjb}
\bibinfo{author}{\bibnamefont{{ A. Cuccoli, A. Fort, A. Rettori, E. Adam, and
  J. Villain}}}, \bibinfo{journal}{Eur. Phys. J. B}
  \textbf{\bibinfo{volume}{12}}, \bibinfo{pages}{39} (\bibinfo{year}{1999}).

\bibitem[{\citenamefont{Alonso and Fern\'andez}(2001)}]{alofer01prl}
\bibinfo{author}{\bibfnamefont{J.~J.} \bibnamefont{Alonso}} \bibnamefont{and}
  \bibinfo{author}{\bibfnamefont{J.~F.} \bibnamefont{Fern\'andez}},
  \bibinfo{journal}{Phys. Rev. Lett.} \textbf{\bibinfo{volume}{87}},
  \bibinfo{pages}{097205} (\bibinfo{year}{2001}).

\bibitem[{\citenamefont{Fern\'andez and Alonso}(2003)}]{feralo03prl}
\bibinfo{author}{\bibfnamefont{J.~F.} \bibnamefont{Fern\'andez}}
  \bibnamefont{and} \bibinfo{author}{\bibfnamefont{J.~J.}
  \bibnamefont{Alonso}}, \bibinfo{journal}{Phys. Rev. Lett.}
  \textbf{\bibinfo{volume}{91}}, \bibinfo{pages}{047202}
  (\bibinfo{year}{2003}).

\bibitem[{\citenamefont{{P. C. E. Stamp and I. S.
  Tupitsyn}}(2004)}]{statup04prb}
\bibinfo{author}{\bibnamefont{{P. C. E. Stamp and I. S. Tupitsyn}}},
  \bibinfo{journal}{Phys. Rev. B} \textbf{\bibinfo{volume}{69}},
  \bibinfo{pages}{014401} (\bibinfo{year}{2004}).

\bibitem[{\citenamefont{{I. S. Tupitsyn, P. C. E. Stamp, and N. V.
  Prokof'ev}}(2004)}]{tupstapro04prb}
\bibinfo{author}{\bibnamefont{{I. S. Tupitsyn, P. C. E. Stamp, and N. V.
  Prokof'ev}}}, \bibinfo{journal}{Phys. Rev. B} \textbf{\bibinfo{volume}{69}},
  \bibinfo{pages}{132406} (\bibinfo{year}{2004}).

\bibitem[{\citenamefont{Fern\'andez and Alonso}(2004)}]{feralo04prb}
\bibinfo{author}{\bibfnamefont{J.~F.} \bibnamefont{Fern\'andez}}
  \bibnamefont{and} \bibinfo{author}{\bibfnamefont{J.~J.}
  \bibnamefont{Alonso}}, \bibinfo{journal}{Phys. Rev. B}
  \textbf{\bibinfo{volume}{69}}, \bibinfo{pages}{024411}
  (\bibinfo{year}{2004}).

\bibitem[{\citenamefont{Fern\'andez and Alonso}(2005)}]{feralo05prb}
\bibinfo{author}{\bibfnamefont{J.~F.} \bibnamefont{Fern\'andez}}
  \bibnamefont{and} \bibinfo{author}{\bibfnamefont{J.~J.}
  \bibnamefont{Alonso}}, \bibinfo{journal}{Phys. Rev. B}
  \textbf{\bibinfo{volume}{72}}, \bibinfo{pages}{094431}
  (\bibinfo{year}{2005}).

\bibitem[{\citenamefont{{W. Wernsdorfer, T. Ohm, C. Sangregorio, R. Sessoli, D.
  Mailly, and C. Paulsen}}(1999)}]{weretal99prl}
\bibinfo{author}{\bibnamefont{{W. Wernsdorfer, T. Ohm, C. Sangregorio, R.
  Sessoli, D. Mailly, and C. Paulsen}}}, \bibinfo{journal}{Phys. Rev. Lett.}
  \textbf{\bibinfo{volume}{82}}, \bibinfo{pages}{3903} (\bibinfo{year}{1999}).

\bibitem[{\citenamefont{Garanin and Chudnovsky}(2009)}]{garchu09prl}
\bibinfo{author}{\bibfnamefont{D.~A.} \bibnamefont{Garanin}} \bibnamefont{and}
  \bibinfo{author}{\bibfnamefont{E.~M.} \bibnamefont{Chudnovsky}},
  \bibinfo{journal}{Phys. Rev. Lett.} \textbf{\bibinfo{volume}{78}},
  \bibinfo{pages}{097206} (\bibinfo{year}{2009}).

\bibitem[{\citenamefont{{Nurit Avraham, Ady Stern, Yoko Suzuki, K. M. Mertes,
  M. P. Sarachik, E. Zeldov, Y. Myasoedov, H. Shtrikman,1 E. M. Rumberger, D.
  N. Hendrickson, N. E. Chakov, and G. Christou}}(2005)}]{avretal05prb}
\bibinfo{author}{\bibnamefont{{Nurit Avraham, Ady Stern, Yoko Suzuki, K. M.
  Mertes, M. P. Sarachik, E. Zeldov, Y. Myasoedov, H. Shtrikman,1 E. M.
  Rumberger, D. N. Hendrickson, N. E. Chakov, and G. Christou}}},
  \bibinfo{journal}{Phys. Rev. B} \textbf{\bibinfo{volume}{72}},
  \bibinfo{pages}{144428} (\bibinfo{year}{2005}).

\bibitem[{\citenamefont{{Y. Suzuki, M. P. Sarachik, E. M. Chudnovsky, S.
  McHugh,R. Gonzalez-Rubio, N. Avraham, Y. Myasoedov, E. Zeldov, H. Shtrikman,
  N. E. Chakov and G. Christou}}(2005)}]{suzetal05prl}
\bibinfo{author}{\bibnamefont{{Y. Suzuki, M. P. Sarachik, E. M. Chudnovsky, S.
  McHugh,R. Gonzalez-Rubio, N. Avraham, Y. Myasoedov, E. Zeldov, H. Shtrikman,
  N. E. Chakov and G. Christou}}}, \bibinfo{journal}{Phys. Rev. Lett.}
  \textbf{\bibinfo{volume}{95}}, \bibinfo{pages}{147201}
  (\bibinfo{year}{2005}).

\bibitem[{\citenamefont{Garanin and Chudnovsky}(2007)}]{garchu07prb}
\bibinfo{author}{\bibfnamefont{D.~A.} \bibnamefont{Garanin}} \bibnamefont{and}
  \bibinfo{author}{\bibfnamefont{E.~M.} \bibnamefont{Chudnovsky}},
  \bibinfo{journal}{Phys. Rev. B} \textbf{\bibinfo{volume}{76}},
  \bibinfo{pages}{054410} (\bibinfo{year}{2007}).

\bibitem[{\citenamefont{Garanin and Chudnovsky}(1997)}]{garchu97prb}
\bibinfo{author}{\bibfnamefont{D.~A.} \bibnamefont{Garanin}} \bibnamefont{and}
  \bibinfo{author}{\bibfnamefont{E.~M.} \bibnamefont{Chudnovsky}},
  \bibinfo{journal}{Phys. Rev. B} \textbf{\bibinfo{volume}{56}},
  \bibinfo{pages}{11102} (\bibinfo{year}{1997}).

\bibitem[{\citenamefont{Garanin}(2008)}]{gar08-DME}
\bibinfo{author}{\bibfnamefont{D.~A.} \bibnamefont{Garanin}},
  \bibinfo{journal}{arXiv:0805.0391}  (\bibinfo{year}{2008}).

\bibitem[{\citenamefont{Chudnovsky et~al.}(2005)\citenamefont{Chudnovsky,
  Garanin, and Schilling}}]{chugarsch05prb}
\bibinfo{author}{\bibfnamefont{E.~M.} \bibnamefont{Chudnovsky}},
  \bibinfo{author}{\bibfnamefont{D.~A.} \bibnamefont{Garanin}},
  \bibnamefont{and}
  \bibinfo{author}{\bibfnamefont{R.}~\bibnamefont{Schilling}},
  \bibinfo{journal}{Phys. Rev. B} \textbf{\bibinfo{volume}{72}},
  \bibinfo{pages}{094426} (\bibinfo{year}{2005}).

\end{thebibliography}

\end{document}